\documentclass{article}

\usepackage{iclr2026_conference,times}

\usepackage[utf8]{inputenc}
\usepackage[T1]{fontenc}
\usepackage{hyperref}
\usepackage{url}
\usepackage{booktabs}
\usepackage{amsfonts}
\usepackage{amsmath}
\usepackage{amssymb}
\usepackage{nicefrac}
\usepackage{microtype}
\usepackage{xcolor}
\usepackage{graphicx}
\usepackage{subcaption}
\usepackage{multirow}
\usepackage{enumitem}

\newcommand{\name}{\textsc{SciCode}}
\newcommand{\ours}{\textsc{SciCode-Verified}}
\newcommand{\nDefects}{262}
\newcommand{\nFound}{263}
\newcommand{\nChanged}{63}
\newcommand{\nTotal}{64}

\title{\ours: How Benchmark Defects Underestimated the\\Scientific-Coding Ability of Language Models}

\author{\normalfont\small
\begin{tabular}[t]{@{}l@{\hspace{1.5em}}l@{}}
\textbf{Sihan Hu} & \textbf{Lyuhan Huang} \\
Hefei National Laboratory & College of Mechanical and Electrical Engineering \\
University of Science and Technology of China & Harbin Engineering University \\
Hefei 230026, China & Harbin 150001, China \\[1.5ex]
\textbf{Youjin Deng}\thanks{Corresponding author: \texttt{yjdeng@ustc.edu.cn}} & \textbf{Kun Chen}\thanks{Corresponding author: \texttt{chenkun@itp.ac.cn}} \\
Hefei National Laboratory & Institute of Theoretical Physics \\
University of Science and Technology of China & Chinese Academy of Sciences \\
Hefei 230026, China & Beijing 100190, China
\end{tabular}%
}

\iclrfinalcopy % non-anonymous arXiv preprint (authors shown)
\begin{document}
\maketitle
\lhead{Preprint} % arXiv preprint header (overrides ``Published as a conference paper at ICLR 2026'')

\begin{abstract}
\name{} is the standard measure of the scientific-coding ability of language models:
research-level problems that demand both frontier scientific theory and its implementation as
working numerical code. It is a component of the Artificial Analysis Intelligence Index and a
standing evaluation in government and national-laboratory suites. Yet its scores have recently
plateaued: the strongest 2026 models cluster tightly around $60\%$ subproblem accuracy, and a
successor model ties its predecessor. We trace this stagnation to defects in the benchmark
itself. A per-problem, domain-expert audit of all 65 test problems uncovers
\nFound{} defects; \textbf{192} of them, spread across \textbf{91\%} of the main
problems, cause correct, instruction-following solutions to be wrongly
rejected---through non-reproducible gold answers, over-tight tolerances, or
self-contradictory specifications. Critically, $78\%$ of these score-suppressing defects
require specialized physics or mathematics knowledge to detect, not mere clerical
proofreading. We corrected every confirmable defect to produce \ours{}. The corrections add
only the specifications a well-posed problem requires, repair grading, and tighten the tests
that were too lenient; every change is recorded with its justification and independently
re-checked by a second domain expert. We re-evaluate twelve frontier model snapshots on the corrected
benchmark and find a substantial recovery: subproblem accuracy rises from $45$--$60\%$ to
$84$--$98\%$, and main-problem accuracy from $9$--$27\%$ to $69$--$92\%$. State-of-the-art
models are far more proficient in scientific coding than \name{} has suggested---the
bottleneck was not model capability, but the quality of the evaluation instrument. We release
\ours{} with its complete audit trail as the corrected public
standard.\footnote{Project repository: \url{https://github.com/flyingwagner/scicode-verified}}
\end{abstract}

\section{Introduction}
\label{sec:intro}

\name{}~\citep{scicode} is the standard instrument for measuring the scientific-coding ability
of language models. Its 80 research-level problems are
drawn from real research across mathematics, physics, chemistry, biology, and materials science,
and each one decomposes a research workflow into cumulative subproblems: the model must absorb
expert-level scientific background, carry a derivation through, choose an appropriate numerical
method, and implement it to research-grade accuracy. As language models are increasingly put to
work on frontier scientific research---literature synthesis, derivation, simulation, and
experiment design---the need for an instrument of exactly this capability grows with it. \name{}
occupies a distinctive niche among widely tracked benchmarks: coding evaluations test software
engineering, science evaluations test question answering, and \name{} demands both at once, on
genuine research problems whose solutions are working numerical code---and adoption has
followed: \name{} is a component of the Artificial Analysis Intelligence Index~\citep{aa-index},
whose public leaderboard scores 280 model configurations against it~\citep{aa-scicode};
government evaluators include it in the UK AI Security Institute's Inspect Evals
suite~\citep{aisi2026inspect}; and national laboratories run it as a standing
evaluation~\citep{lbnl2026cborg}. Recent model cards and technical reports likewise increasingly
evaluate on \name{} alongside established
benchmarks~\citep{gemini31procard,kimi2026k25,glm2025arc,openai2026gpt55,seed21,aa-musespark}.

For an instrument of this standing, however, its scores behave strangely. Most tellingly, the
frontier is numerically indistinguishable: on the public leaderboard, the strongest 2026 models
sit within four points of each other---Claude Fable 5, the
strongest public model of
mid-2026 on the overall Intelligence Index, scores $60\%$;
Gemini 3.1 Pro and Kimi K3 score $59\%$; GPT-5.6 Sol ties its predecessor GPT-5.5 at
$56\%$~\citep{aa-scicode}.\footnote{Artificial Analysis \name{} leaderboard~\citep{aa-scicode},
accessed 2026-07-21: Claude Fable 5 $60\%$, Gemini 3.1 Pro $59\%$, Kimi K3 $59\%$, Muse Spark 1.1
$58\%$; the next nine frontier entries score $53$--$56\%$, among them GPT-5.5 and its successor
GPT-5.6 Sol, tied at $56\%$; the 280 listed configurations (multiple settings per model) averaged
$42.9\%$ in the 2026-06-02 snapshot.} An instrument that cannot tell a flagship from
its predecessor leaves only two explanations for its
low scores: either scientific coding genuinely lies beyond today's models, or the
instrument itself is miscalibrated.

\begin{figure}[t]
\centering
\includegraphics[width=\textwidth]{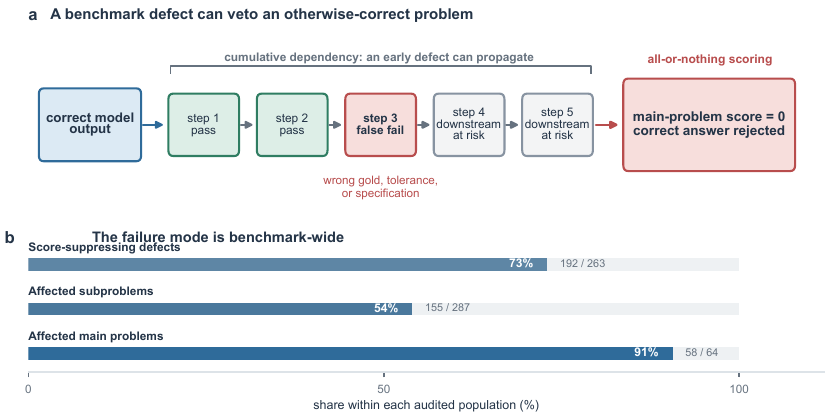}
\caption{How score-suppressing defects become a benchmark-wide failure mode. \textbf{(a)}
Because \name{} subproblems are cumulative, an early defect can propagate downstream; because
main-problem scoring is all-or-nothing, one false failure can veto an otherwise-correct problem.
\textbf{(b)} Of the \nFound{} defects identified, 192 (73\%) reject correct solutions; these touch 155 of 287
scored subproblems (54\%) across 58 of 64 main problems (91\%).}
\label{fig:compression}
\end{figure}

We find that much of the gap stems from the benchmark's design and its defects, not from the
models. Our domain-expert review of all 65 test problems finds \nFound{} defects. We fix all of them but one\footnote{The one we cannot fix is problem~2
(\texttt{Gaussian\_Beam\_Focus}) itself: its specification
fixes no unique answer and so admits no verifiable gold, so we drop the problem rather than
ship an arbitrary gold (\S\ref{sec:background}).} to
produce \ours{}, the corrected benchmark released with this paper. Two properties of \name{}'s
design make defects costly here: subproblems form a cumulative chain---step $k$ builds on steps
$1{:}k{-}1$---so a defective early step contaminates every step downstream; and scoring is
all-or-nothing, so one defective subproblem forfeits the main problem. The confirmed defects are dense and mostly scientific: \textbf{192}
cause a correct, instruction-following solution to be graded wrong---through non-reproducible or
incorrect gold answers, over-tight tolerances, over-specified randomness, or incomplete and
self-contradictory specifications---and touch \textbf{91\%} of the main problems (58 of
\nTotal{}), while \textbf{150 of the 192} (78\%) are recognizable as defects only with specialist
knowledge of the relevant physics or mathematics. Since every model meets the same broken steps, a defect shared by all models becomes a
ceiling shared by all models---which is precisely the compression and stagnation observed on the
leaderboard. The corrections behind \ours{} stay minimal: each either supplies a constraint that
a well-posed problem requires but the prompt left implicit, or repairs grading, and where a test was too
lenient we tightened it, so no corrected problem admits a wrong answer (\S\ref{sec:taxonomy}).
Every change is recorded with its justification and independently re-checked (\S\ref{sec:gate}).

Fixing \name{} is hard for a specific reason: most of its defects are scientific rather than
clerical---principle-level gaps in the specification as much as numerical errors in gold values
and tolerances---so both recognizing and correcting them demands domain specialists. A wrong gold value or tolerance need not fail any surface check, and
recognizing it---let alone computing the right one---means redoing the physics, problem by
problem, across five research fields. Our audit therefore worked at the level of the science
rather than the code: an independent reviewer re-derived each problem's targets and stress-tested
its specification, and a second, adversarial pass re-verified every flagged defect to eliminate
false alarms. Approved changes are recorded with their justification, and each release is mechanically checked against that record (\S\ref{sec:gate}). The result, \ours{}, can now serve \name{}'s original purpose: to
judge models' true command of frontier scientific theory and of the code that implements it.

\paragraph{Contributions.}
\begin{itemize}[leftmargin=1.2em,itemsep=2pt,topsep=2pt]
  \item \textbf{The corrected benchmark \ours{}} (\S\ref{sec:gate}): a corrected release of the
    \nTotal{} verifiable test problems, produced so that every correction can itself be
    audited---each change is recorded with its reason, and the released files are regenerated
    from that record and mechanically checked against it.
  \item \textbf{A defect taxonomy for \name{}} (\S\ref{sec:taxonomy}): we manually verify
    \nDefects{} defects across \nChanged{} of the \nTotal{} test problems---each grounded in the
    actual original-vs-corrected change and re-checked by an independent adversarial pass---and sort
    every one along two axes: its \emph{nature} (\textsc{scientific}, needing physics/mathematics
    judgment to detect, vs \textsc{surface}, a typo, broken cross-reference, or code-interface slip)
    and its \emph{direction} (rejects a correct solution, hardens a too-lenient test, or neutral
    cleanup). The resulting cross-tabulation (Table~\ref{tab:tax}) makes the expertise-gated
    structure of the score suppression legible.
  \item \textbf{A matched before/after re-evaluation of twelve frontier model snapshots}
    (\S\ref{sec:impact}): correction raises subproblem accuracy from the
    $45$--$60\%$ they score on the original benchmark to $84$--$98\%$ and main-problem accuracy to
    $69$--$92\%$ (a $29$--$40$ point subproblem gain per model), and can change their relative
    ranking (single pass@1 runs)---isolating the benchmark's own contribution
    to the measured score.
\end{itemize}

\section{Background: \name{} and its evaluation protocol}
\label{sec:background}

\name{}~\citep{scicode} contains 80 main problems, split into 15 development and
65 test main problems. The 15 development problems ship public
ground-truth code for development and few-shot use, so measurement---ours, as in all published
results---uses the 65-problem test split. Each subproblem
ships a Python function header, a docstring specifying inputs/outputs, optional expert-written
background, and domain test cases the candidate solution must pass. Subproblems are solved \emph{cumulatively}: step $k$ may call the gold (reference)
functions of steps $1{:}k{-}1$. A main problem counts as solved only if \emph{all} its subproblems
pass---an all-or-nothing rule under which a single bad gold value or over-tight tolerance fails the
entire problem. We report this as \emph{main-problem} accuracy, and use \emph{step} and
\emph{subproblem} interchangeably.

A domain test case can wrongly fail a correct solution in two ways. First, the precomputed gold
answer it
compares against may itself be wrong, or reproducible by no method---so any correct output disagrees
with it. Second, the comparison (a numerical \texttt{allclose}) may be too strict: a tolerance
tighter than the method warrants, or one that silently demands a convention the prompt never fixed
(a sign, a unit, an index origin, a grid, or a random-number order). Either way, a correct,
instruction-following solution is liable to be graded wrong.

We audit all 65 test problems. One of them, problem~2
(\texttt{Gaussian\_Beam\_Focus}), is structurally underdetermined: its specification fixes no
unique answer, and its gold is reproducible by no method. Unlike every other defect we encounter,
it cannot be corrected into a verifiable problem, so we drop it rather than ship an arbitrary
gold. (Published 65-problem scores therefore already include one problem that
no correct solution can pass.)

\section{Re-evaluation on \ours{}}
\label{sec:impact}

\begin{figure}[t]
\centering
\includegraphics[width=\textwidth]{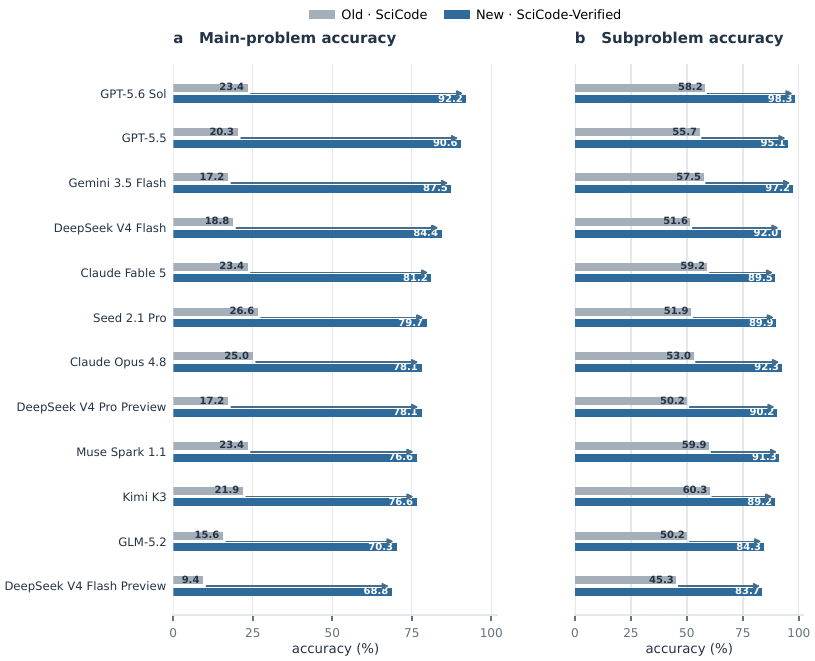}
\caption{Matched before/after re-evaluation of twelve frontier model snapshots (with background, pass@1,
same harness throughout---only the benchmark data differs; Table~\ref{tab:beforeafter}).
Within each model, grey and blue bars show accuracy on original \name{} and \ours{},
respectively; the arrow marks the matched old$\rightarrow$new change. Correction lifts every
model by $53$--$70$ main-problem points \textbf{(a)} and $29$--$40$
subproblem points \textbf{(b)}. It also restores discrimination: GPT-5.5 and DeepSeek V4 Pro Preview move
from two main problems apart to eight, while Seed 2.1 Pro leads the original main-problem
score yet ranks sixth on \ours{}. The two DeepSeek V4 Preview rows on \ours{} are the mean of
three runs; all other points are single pass@1 runs.}
\label{fig:beforeafter}
\end{figure}

\paragraph{Frontier models on \ours{}.} We evaluate twelve frontier model snapshots, listed in
Table~\ref{tab:beforeafter}, on both the original benchmark and \ours{} under a matched
with-background harness; only the benchmark data differ.
On the original benchmark, they score \textbf{45--60\%} on subproblems and \textbf{9--27\%} on main
problems. On \ours{}, the ranges rise to \textbf{84--98\%} and \textbf{69--92\%}, respectively
(Table~\ref{tab:beforeafter}; Figure~\ref{fig:beforeafter}). Every model gains $29$--$40$ subproblem
points and $53$--$70$ main-problem points. Correcting subproblem defects distributed across most
problems removes false failures that previously caused entire main problems to be scored as
incorrect, producing the larger gains in main-problem accuracy. We analyze the prevalence and
effects of these defects in \S\ref{sec:taxonomy}.

At the upper end, the corrected benchmark approaches saturation: GPT-5.6~Sol reaches $98.3\%$
subproblem and $92.2\%$ main-problem accuracy. Across the full model set, however, correction does
not shift all scores uniformly: it both widens previously compressed gaps and changes model
rankings. On the original benchmark, GPT-5.5 passes only two more main problems than
DeepSeek~V4~Pro~Preview; on \ours{}, it passes eight more. Rankings also change: Seed~2.1~Pro ranks
first by original main-problem accuracy but sixth on \ours{}. The full pattern is shown in
Figure~\ref{fig:beforeafter} and Table~\ref{tab:beforeafter}.

Following the Artificial Analysis leaderboard~\citep{aa-scicode}, we report pass@1. Except for the
two corrected-benchmark DeepSeek~V4~Preview rows, which average three runs, each entry is a single
run. Pass@1 scores vary across runs, so small differences between models, particularly in
main-problem accuracy, should not be over-interpreted.

\begin{table}[t]
\centering
\small
\caption{Frontier models on the original vs.\ corrected benchmark under the \emph{same} harness
(with background, pass@1, two-environment OR grading; \S\ref{sec:eval-detail})---only the
benchmark data differs. \emph{Original} columns are our re-run of the original upstream release on
the same 64 problems. \ours{} DeepSeek V4 Preview rows ($^\ast$) are the mean of three runs on an earlier harness revision. Every model
gains $29$--$40$ subproblem points after correction.}
\label{tab:beforeafter}
\begin{tabular}{lcccc}
\toprule
 & \multicolumn{2}{c}{Original} & \multicolumn{2}{c}{\ours{}} \\
\cmidrule(lr){2-3}\cmidrule(lr){4-5}
Model & sub.\,\% & main\,\% & sub.\,\% & main\,\% \\
\midrule
GPT-5.6 Sol        & 58.2 & 23.4 & 98.3 & 92.2 \\
GPT-5.5            & 55.7 & 20.3 & 95.1 & 90.6 \\
Gemini 3.5 Flash   & 57.5 & 17.2 & 97.2 & 87.5 \\
DeepSeek V4 Flash  & 51.6 & 18.8 & 92.0 & 84.4 \\
Claude Fable 5     & 59.2 & 23.4 & 89.5 & 81.2 \\
Seed 2.1 Pro       & 51.9 & 26.6 & 89.9 & 79.7 \\
Claude Opus 4.8    & 53.0 & 25.0 & 92.3 & 78.1 \\
DeepSeek V4 Pro Preview & 50.2 & 17.2 & 90.2$^\ast$ & 78.1$^\ast$ \\
Muse Spark 1.1     & 59.9 & 23.4 & 91.3 & 76.6 \\
Kimi K3            & 60.3 & 21.9 & 89.2 & 76.6 \\
GLM-5.2            & 50.2 & 15.6 & 84.3 & 70.3 \\
DeepSeek V4 Flash Preview & 45.3 &  9.4 & 83.7$^\ast$ & 68.8$^\ast$ \\
\bottomrule
\end{tabular}
\end{table}

\paragraph{Without the background.} \name{}'s most realistic setting withholds the
expert-written background. With it, GPT-5.5, Gemini 3.5 Flash, and Claude Opus 4.8 solve
\textbf{92--97\%} of subproblems on \ours{}, approaching saturation on this metric. Without it,
their subproblem accuracy falls to \textbf{80--83\%} and their main-problem accuracy to
\textbf{52--66\%} (Table~\ref{tab:nobg}). Thus, while the with-background benchmark leaves little
room on subproblem accuracy, the no-background setting retains meaningful headroom. We analyze
the fail-without/pass-with behavior of these three models in \S\ref{sec:bg-knowledge}.

\begin{table}[t]
\centering
\small
\caption{With- vs.\ no-background accuracy on \ours{} (\%, with-background grading as in
Table~\ref{tab:beforeafter}; two-environment OR). Withholding the expert-written
background---\name{}'s most realistic setting---costs $10$--$22$ subproblem and $20$--$36$
main-problem points, yet GPT-5.5, Gemini 3.5 Flash, and Claude Opus 4.8 still clear
$80$--$83\%$ of subproblems. \textsc{GLM-5.2}
has no no-background run; DeepSeek V4 Preview rows are the mean of three runs.}
\label{tab:nobg}
\begin{tabular}{lcccc}
\toprule
 & \multicolumn{2}{c}{Subproblem\,\%} & \multicolumn{2}{c}{Main-problem\,\%} \\
\cmidrule(lr){2-3}\cmidrule(lr){4-5}
Model & bg & no-bg & bg & no-bg \\
\midrule
GPT-5.5            & 95.1 & 82.6 & 90.6 & 65.6 \\
Gemini 3.5 Flash   & 97.2 & 80.5 & 87.5 & 51.6 \\
Claude Opus 4.8    & 92.3 & 81.9 & 78.1 & 57.8 \\
DeepSeek V4 Pro Preview   & 90.2 & 68.1 & 78.1 & 46.4 \\
DeepSeek V4 Flash Preview & 83.7 & 62.8 & 68.8 & 40.6 \\
\bottomrule
\end{tabular}
\end{table}

\subsection{What knowledge the background supplies}
\label{sec:bg-knowledge}

To understand what the background contributes, we analyze 31 subproblems that at least two of
three models---GPT-5.5, Gemini 3.5 Flash, and Claude Opus 4.8---fail without the background but
pass with it. To identify root causes, we compare the paired generations, reproduce the
computations, and trace downstream failures to their originating step. Among model-attributable
failures, we identify four recurring root causes.

\emph{(i) Reassembling a multi-step derivation.} In \#32 (optically bound nanoparticle arrays),
the background walks from the optical-binding force to the phonon Hamiltonian---trap stiffness
$k_i=\alpha E_i^2/w^2$, binding-shifted frequencies $\Omega_i=\sqrt{(k_i+\sum_{j}k_{ij})/m}$,
signed hopping $g_{ij}=-k_{ij}/(2m\sqrt{\Omega_i\Omega_j})$. Without it, the three models produce
three \emph{mutually inconsistent} Hamiltonians: Claude Opus 4.8 omits the optical-trap stiffness,
GPT-5.5 omits the binding-induced frequency shift, and Gemini 3.5 Flash doubles the trap
stiffness. Here, the background provides derivational scaffolding that the models do not reliably
reconstruct on their own.

\emph{(ii) Textbook priors overriding the stated formulation.} In \#54 (SUPG finite elements),
the problem statement explicitly specifies a non-standard weak form, yet all three models silently
replace it with the \emph{textbook} SUPG/Nitsche formulation. Restating the equations in the
background at the point of implementation keeps the models on the stated variant, suggesting that
a strong textbook prior overrides information already present in the prompt.

\emph{(iii) Specialized domain facts.} In \#69 (Raman response of a layered electron gas), the
models get a domain-specific fact wrong: the backscattering photon momentum transfer is $2k$,
not $k$.

\emph{(iv) Near-miss implementation errors.} In \#79 (Nos\'e--Hoover chains), the models use the
correct reversible-integrator factorization but make isolated bookkeeping errors, such as halving
a half-step twice or reusing a stale force array. Here, the background acts as an implementation
check rather than supplying a missing method.

Consistent with these being knowledge gaps rather than noise, the no-background penalty is larger
for the DeepSeek V4 Preview pair (Table~\ref{tab:nobg}): they lose $21$--$22$ subproblem points,
compared with $10$--$17$ for the three models analyzed in this diagnostic.

\paragraph{Library-version-robust grading.}
\label{sec:eval-detail}
We grade each saved solution in two pinned NumPy/SciPy environments and count it as correct if it
passes either. This recovers $7$--$9$ of 287 subproblems per model that otherwise fail only because
of library-version incompatibilities; the rescued sets are nearly identical across models
(Appendix~\ref{app:orgain}). The rule therefore reduces environment-induced false negatives
without favoring any model.

\section{What we corrected: a defect taxonomy}
\label{sec:taxonomy}

We examined and cleaned the test problems one at a time, then added a round of
independent adversarial review, to produce \ours{}. We sort each defect into two broad
kinds---\textsc{scientific}, whose identification requires relevant physics or mathematics knowledge,
and \textsc{surface}, a corpus, code, or text issue---and by \emph{direction}: whether the
original was \emph{too strict or wrong} (failing a correct solution), \emph{too lenient} (a weak
test we tightened), or \emph{neither} (a neutral cleanup). Of the \nDefects{} defects, \textbf{177
are \textsc{scientific}} and \textbf{85 are \textsc{surface}}. By direction, \textbf{192 of
\nDefects{}} defects ($\textbf{73\%}$) are too strict or wrong and therefore suppress measured model
accuracy; \textbf{8} are too lenient and \textbf{62} are neutral. Of these 192 score-suppressing
defects, \textbf{150} ($\textbf{78\%}$) are scientific. They span \textbf{155 of 287} scored
subproblems across \textbf{58 of 64} problems (Table~\ref{tab:tax}).
Across the four models included in our subproblem-level attribution analysis, all 164 newly passed
subproblems trace to a documented correction.

\begin{table}[t]
\centering
\small
\caption{Defect taxonomy of the \name{} audit: \nDefects{} defects in \nChanged{} of \nTotal{} test
problems, corrected in \ours{}, by \emph{nature} $\times$ \emph{direction}. Two-thirds (177) are \textsc{scientific}
defects that require physics/mathematics judgment to detect; and of the 192 that reject a correct
solution, 150 ($78\%$) are scientific. The only ``too-lenient'' entries are weak \emph{tests}
we \emph{tightened} ($^{\uparrow}$: stricter, not easier). Provenance: identified in expert review
\textbf{109}, recovered by the original$\to$corrected diff and confirmed by expert review \textbf{153}.}
\label{tab:tax}
\begin{tabular}{lrrrr}
\toprule
 & \multicolumn{3}{c}{Direction (per defect)} & \\
\cmidrule(lr){2-4}
Defect nature & Strict/wrong & Neutral & Weak-test$^{\uparrow}$ & Total \\
\midrule
\textsc{Scientific} (needs physics/math judgment to detect) & 150 & 21 & 6 & 177 \\
\textsc{Surface} (corpus / code / text, trivial)           & 42  & 41 & 2 & 85 \\
\midrule
\textbf{Total} & \textbf{192} & \textbf{62} & \textbf{8} & \textbf{262} \\
\bottomrule
\end{tabular}
\end{table}

\paragraph{Scientific failure mechanisms.} The 177 scientific defects fall into seven subtypes
(Table~\ref{tab:cat5}). The largest is unspecified conventions (77), where the gold fixes a
choice that the prompt leaves free, followed by wrong gold (33) and RNG-dependent grading (22).
The seven examples below show how each subtype distorts evaluation.

\begin{table}[t]
\centering
\small
\caption{Sub-structure of the two defect natures (\nDefects{} defects). \textsc{Scientific}
subtypes require domain judgment to detect; \textsc{surface} subtypes do not. A full per-defect
map is in the appendix.}
\label{tab:cat5}
\begin{tabular}{lrl}
\toprule
Subtype & Count & Example \\
\midrule
\multicolumn{3}{@{}l}{\textit{\textsc{Scientific} (177) --- needs physics/math judgment to detect}}\\
Unspecified convention                 & 77 & \#74 Householder sign sector; index base \\
Wrong gold (method / sign / symmetry)  & 33 & \#22 rotation sign; \#13 $A_z$ parity \\
RNG-dependent grading                 & 22 & \#13/\#46/\#50 seeded/exact-scheme $\to$ statistical \\
Non-discriminating test (physics)      & 14 & \#5.1 palindromic Lanczos input \\
Over-tight tolerance                   & 13 & \#30.3 target is a roundoff fingerprint \\
Spec$\leftrightarrow$gold contradiction & 9 & \#8.1 strict boundary vs gold's $\geq$ \\
Invalid / non-reproducible target      & 9  & \#28.1 reproducible by no method \\
\midrule
\multicolumn{3}{@{}l}{\textit{\textsc{Surface} (85) --- corpus / code / text, trivial}}\\
Interface / return-contract mismatch   & 29 & \#65 \texttt{def tensor()} vs variadic doc \\
Truncation / garble / dropped constant & 20 & \#15 dropped $\hbar$ mantissa \\
Broken cross-reference                 & 16 & dangling ``step \_\_\_'' pointers \\
Trivially broken test                  & 12 & \#66.6 misplaced paren, one-sided tol. \\
Typo / formatting                      & 8  & ``fcous'' $\to$ ``focus'' \\
\bottomrule
\end{tabular}
\end{table}

\begin{itemize}[leftmargin=1.2em,itemsep=2pt,topsep=2pt]
  \item \emph{Unspecified convention, \#74.} Householder QR~\citep{householder1958}
    successively applies reflectors to eliminate entries below the diagonal. In the final column,
    a $1\times1$ reflector eliminates nothing: applying it only flips the sign of the last column
    of $Q$ and the last diagonal entry of $R$, leaving the product $QR$ unchanged. Applying this
    reflector and stopping before it are therefore both valid conventions. The prompt did not
    specify which to use, while the gold accepted only the former; we now state that convention
    explicitly.
  \item \emph{Wrong gold (sign error), \#22.} The rotation-coefficient recurrence~\citep{gumerov2004}
    shipped a sign error ($-2\!\to\!+2$) that makes the transfer matrix non-unitary. The corrected
    sign reproduces the unique unitary Wigner coefficient to $10^{-14}$.
  \item \emph{RNG-dependent grading, \#68.} The original diffusion Monte-Carlo
    test~\citep{metropolis1953,reynolds1982} depended on a particular random-number sequence.
    We instead check the final statistic, requiring the mean energy to be within $0.05$\,Ha of the
    helium ground-state energy.
  \item \emph{Non-discriminating test, \#5.1.} The first Lanczos test~\citep{lanczos1950}
    paired a reflection-symmetric tridiagonal matrix with a palindromic starting vector. Because the
    matrix preserves reflection symmetry, all Krylov vectors remain in the four-dimensional symmetric
    subspace, although the test requests six Lanczos vectors. The iteration must therefore break down
    before exercising all requested steps. We replace the starting vector with an asymmetric one; its
    first six Krylov vectors are linearly independent, so the same test now exercises the full run.
  \item \emph{Over-tight tolerance, \#30/\#67.} For the helium Slater--Jastrow
    wavefunction~\citep{jastrow1955} and the layered electron gas~\citep{jain1985}, an
    \texttt{allclose} at $\texttt{rtol}=10^{-15}$ (below
    double precision) or against a rounding fingerprint fails every correct implementation. We
    loosen minimally to admit the legitimate method/rounding choice while still rejecting wrong
    answers.
  \item \emph{Spec$\leftrightarrow$gold contradiction, \#8.} For Fourier-plane spatial
    filtering~\citep{goodman1996}, the prompt requires a strict high-pass boundary, but the gold
    uses an inclusive ($\geq$) boundary. We regenerate the gold to match the prompt.
  \item \emph{Invalid or non-reproducible target, \#28.1.} For Gaussian-beam propagation through
    a lens system~\citep{kogelnikli1966}, the target is reproducible by no standard method and
    contradicts a later subproblem. We regenerate it with the prompt-prescribed method.
\end{itemize}

\paragraph{The corrections preserve difficulty.} To address these defects, we added necessary
conventions, corrected erroneous gold answers, and adopted more appropriate verification criteria.
These changes prevent valid solutions from being misgraded, allowing models to achieve higher scores
without reducing the scientific difficulty of the problems.

\section{Related work}
\label{sec:related}

Benchmark-verification work has addressed several kinds of defects. \textsc{UTBoost}~\citep{utboost}
strengthens overly permissive \textsc{SWE-bench}~\citep{swebench} tests against erroneous patches,
while \textsc{SWE-bench Verified}~\citep{swebenchverified} human-validates a 500-task subset. Other
work corrects defects that suppress accuracy: label errors in factual-consistency
benchmarks~\citep{arellmsbetter}, wrong golds in \textsc{HumanEval}~\citep{evalplus}, and defects in
data-engineering tasks~\citep{eltbenchverified}. \textsc{MMLU-Redux}~\citep{mmluredux} and Platinum
Benchmarks~\citep{platinum} further show that residual label noise matters near saturation. \name{}
contains both kinds of problems: some tests are too weak, while incorrect gold values, over-tight
tolerances, and unspecified numerical conventions reject valid solutions.

SciCode's issue tracker documents similar grading defects, including over-strict
tolerances\footnote{\url{https://github.com/scicode-bench/SciCode/issues/8}} and gold solutions that
fail their own tests.\footnote{\url{https://github.com/scicode-bench/SciCode/issues/43}}

\section{Limitations}
\label{sec:limitations}

The development split releases gold answers and ground-truth code and is therefore not used for
model evaluation. Our audit and corrections cover only the 65-problem test split used for evaluation;
all 263 identified defects come from this split. Pass@1 scores exhibit run-to-run variation. Because
many models are already near saturation on \ours{}, small score differences among them are more
sensitive to this variation. Our reported subproblem accuracies on the original benchmark also differ
somewhat from those on the Artificial Analysis leaderboard~\citep{aa-scicode}.
Although the audit underwent multiple rounds of review, all reviews were conducted by the authors
and therefore do not constitute external blind validation.

\section{Conclusion}
\name{}'s ability to distinguish frontier models has been limited by defects in the benchmark
itself. Our audit of all 65 test problems identifies \nFound{} defects; we correct every repairable
defect and exclude one problem that cannot be made verifiable. Re-evaluating twelve frontier
model snapshots on \ours{}, we find that subproblem accuracy rises from $45$--$60\%$ to $84$--$98\%$ and
main-problem accuracy from $9$--$27\%$ to $69$--$92\%$. We release the corrected benchmark
together with a complete record of its defects and corrections to support fair and reproducible
evaluation of scientific-coding ability.

\ours{} evaluates single-problem scientific coding rather than long-horizon agentic workflows
involving tool use, environment interaction, and iterative development. Developing reliable
benchmarks for such workflows remains an important direction for future work.

\subsubsection*{Reproducibility statement}
Dataset, harness, and audit trail are public at
\url{https://github.com/flyingwagner/scicode-verified}. The \texttt{data} release contains the
corrected benchmark (\texttt{problems\_test.jsonl}, \texttt{test\_data\_cleaned.h5}) together with
\texttt{manifest.json}, whose md5 checksums the harness re-verifies at startup so results cannot
silently be produced from stale data. The repository holds the evaluation harness
(\texttt{eval\_clean/}: official \name{} prompt templates, cumulative scoring, two-environment OR
grading), the per-round record of every change (\texttt{ledger/}), the release gate, and the analysis
files behind the statistics in this paper (per-subproblem defect list, before/after flip sets,
fixed-output re-grading summary, per-environment grading gains). Original-benchmark runs use the identical harness with the original upstream
text and \texttt{h5} as input (md5-pinned). All scores are pass@1 collected in
June--July 2026; all runs are single-sample except the \ours{} DeepSeek V4 Preview rows, which average three.
API models were queried with reasoning enabled and provider-default sampling. Failed API
calls were retried and per-step outputs and grading verdicts are cached.

\subsubsection*{Acknowledgments}
We thank Xiansheng Cai, Shuo Chen, Wenbo Shen, Siheng Chen, and Linfeng Zhang for helpful
discussions.
K.C. is supported by the Strategic Priority Research Program of the Chinese Academy of Sciences
under Grant No.~XDB1680102, the National Key Research and Development Program of China under
Grant No.~2024YFA1408604, and the National Natural Science Foundation of China under Grants
No.~12474245 and No.~12447103. Y.D. and S.H. are supported by the National Natural Science
Foundation of China under Grant No.~12275263, the Quantum Science and
Technology---National Science and Technology Major Project under Grant
No.~2021ZD0301900, and the Natural Science Foundation of Fujian Province of China under Grant
No.~2023J02032.

\bibliographystyle{iclr2026_conference}
\bibliography{refs}

\newpage
\appendix

\section{Audit and correction workflow}
\label{sec:gate}

We went through the test problems one at a time. For each, we ran its reference solution against its
own test cases, marked every defect we found---a wrong or non-reproducible gold, an over-tight
tolerance, RNG-dependent grading, an unspecified convention, or a truncated
or self-contradictory prompt---and applied a targeted fix, regenerating the affected gold targets
when a text edit alone could not resolve it. Every change is logged, every corrected reference
solution passes its own (corrected) tests before release, and the released dataset and evaluation
files are rebuilt directly from these per-problem corrections.

The audit ran as successive review rounds, each structured as audit $\to$ fix $\to$ adversarial
re-review $\to$ release gate. Each approved change is recorded (problem, field, before,
after, rationale, round) and applied to a single per-problem source file, from which the released
files are regenerated. Before each release, an automated check verifies both directions---every
recorded change appears in the release, and every changed field is backed by a record---and re-runs
the gold self-tests. Every fix then faced an independent adversarial re-review that had to reproduce the
defect and attempt to refute the fix. Where it could not confirm, fixes were overturned and
redone. Problem 13 is a documented example: a parity fix triggered a re-audit and a test
redesign spanning six downstream subproblems.

Two disclosures. First, regenerating the gold targets changed some precomputed values: 69
target groups differ from the upstream release. 64 of these correspond to recorded changes;
the remaining five (subproblems 1.1, 61.1, 69.5, 73.1, 80.6) changed only because byte-identical
gold code yields different values in a modern software environment. On four of those five, all
four models with matched before/after runs (\S\ref{sec:impact}) flip from fail to pass; the upstream precomputed values had become unreachable.
These five are \emph{not} counted among the \nDefects{} defects. Second, each problem was audited
by one domain expert and every approved fix re-checked by a second. Because that second pass
confirms rather than independently annotates, we report no inter-annotator agreement
(\S\ref{sec:limitations}).

\section{Two-environment OR grading: measured effect}
\label{app:orgain}

Grading runs each step's tests under two pinned environments---\texttt{2024} (numpy 1.26/scipy 1.13
era) and \texttt{2025} (current)---and a step passes if it passes in either
(\S\ref{sec:eval-detail}). Table~\ref{tab:orgain} reports, per model, the steps whose verdict the
OR rule changes relative to the \texttt{2024} stack alone, extracted from the released per-step
grading caches (\texttt{analysis/or\_gain.json}).

\begin{table}[h]
\centering
\small
\begin{tabular}{lccc}
\toprule
Model & Steps rescued (of 287) & Step ids & Main problems flipped \\
\midrule
GPT-5.5           & 8 & 65.6, 80.1--80.7        & 2 (65, 80) \\
Gemini 3.5 Flash  & 7 & 65.6, 80.1--80.6        & 0 \\
Claude Opus 4.8   & 9 & 60.5, 65.5--65.6, 80.1--80.6 & 2 (60, 65) \\
GLM-5.2           & 8 & 65.5--65.6, 80.1--80.6  & 1 (65) \\
\bottomrule
\end{tabular}
\caption{Steps that pass only under the \texttt{2025} grading stack. The rescued sets are nearly
identical across models, so the OR rule favors no model.}
\label{tab:orgain}
\end{table}

Every diagnosed rescue is an environment artifact, not a correctness difference. (i)~The
\texttt{2024} stack lacks \texttt{matplotlib}. Problem 80's shared dependency header does
\texttt{from mpl\_toolkits.mplot3d import Axes3D}, so every step of that problem fails at import
in that stack regardless of the submitted code; this accounts for 6--7 of each model's rescues. (ii)~In problem
65, scipy's \texttt{sqrtm} emits extended-precision (\texttt{complex256}) arrays for
near-singular density matrices, which the older numpy's \texttt{linalg} rejects with a
\texttt{TypeError}; this accounts for 1--2 rescues per model, identically across independently written solutions.
(iii)~One rescue (Claude Opus 4.8, step 60.5, a large Monte Carlo) is verified from the cache but
was not diagnosed within our reproduction time budget; we report it as undiagnosed. Two
caveats. First, the grader tries the \texttt{2024} stack first and short-circuits on a pass, so the
caches exhibit only the \texttt{2024}-fail\,$\to$\,\texttt{2025}-pass direction (re-running a
passing step in the second stack could not change an OR verdict). Second, main-problem effects
differ across models only because of independent failures elsewhere in the same problems:
Gemini~3.5~Flash, for example, gains no main problem despite seven rescued steps.

\section{Fixed-output re-grading: grading layer vs.\ specification layer}
\label{app:regrade}

To separate how much of the before$\to$after recovery comes from correcting the \emph{grading}
(tests, gold targets) versus correcting what the model \emph{sees} (specifications), we re-graded
every model's original-run generations, byte-unchanged, against the corrected tests and gold
targets, with the same harness, the same two-environment OR, and the same per-step time cap. The
officially skipped subproblems keep their injected gold code. No new model calls are involved.

\begin{table}[h]
\centering
\small
\begin{tabular}{lcccc}
\toprule
Model & Original & Re-graded outputs & Corrected (fresh gen) & Grading-layer share \\
\midrule
GPT-5.5          & 160 · 13 & 200 · 25 & 273 · 58 & 40/113 (35\%) \\
Gemini 3.5 Flash & 165 · 11 & 196 · 24 & 279 · 56 & 31/114 (27\%) \\
Claude Opus 4.8  & 152 · 16 & 175 · 20 & 265 · 50 & 23/113 (20\%) \\
GLM-5.2          & 144 · 10 & 168 · 19 & 242 · 45 & 24/98 (24\%) \\
DeepSeek V4 Pro Preview   & 144 · 11 & 167 · 21 & --- & --- \\
DeepSeek V4 Flash Preview & 130 · 6  & 160 · 19 & --- & --- \\
\bottomrule
\end{tabular}
\caption{Cells are subproblems/287~·~main problems/64 (with background). ``Re-graded outputs''
scores the \emph{original} generations under the corrected grading. The fresh-generation column
is omitted for the DeepSeek V4 Preview pair, whose corrected-benchmark runs used an earlier harness revision
(starred in Table~\ref{tab:beforeafter}).}
\label{tab:regrade}
\end{table}

Re-grading alone recovers $20$--$35\%$ of each model's step-level gain, but nearly doubles or
triples main-problem counts for the weaker models (DeepSeek~V4~Flash~Preview $6\to19$). Under
all-or-nothing scoring, a single mis-graded subproblem vetoes an otherwise-correct problem. Three
caveats. (i)~Sixteen scored subproblems changed function or return signatures during correction;
old code graded against them produces artifact failures (18 pass$\to$fail cells across all
models). Excluding all sixteen, the grading-layer gains on the 271 signature-stable subproblems
are $+24$ to $+42$ per model. The table \emph{understates} the grading-layer effect; it does not
inflate it. (ii)~A handful of pass$\to$fail flips per model ($\leq 6$; e.g.\ subproblems 8.1,
9.1, 31.1) are legitimate. The corrected gold targets encode conventions the original
generations could not have known, so the corrected grading is stricter there, consistent with
corrections that add missing constraints. (iii)~The complementary ``specification-layer'' share
conflates the fairer prompts with fresh sampling; splitting those would require paid
regeneration ablations, which we did not run.

% AUTO-GENERATED by build_appendix.py from defects.json — do not hand-edit.
\section{Complete per-problem defect log}\label{app:log}
This appendix lists every defect in \ours{}, grouped by problem and derived from the released decision ledger and the original$\rightarrow$corrected diff. Each entry gives the subproblem, defect \emph{nature}:\,\emph{subtype} (\textsc{scientific}=needs physics/math judgment to detect, \textsc{surface}=corpus/code/text; subtypes are named as in Table~\ref{tab:cat5}), direction, and a one-line scientific reason for the correction. Direction labels correspond to the columns of Table~\ref{tab:tax}: ``too-strict/wrong'' $\leftrightarrow$ Strict/wrong, ``neutral-cleanup'' $\leftrightarrow$ Neutral, ``too-lenient (test tightened)'' $\leftrightarrow$ Weak-test$^{\uparrow}$. The ledger also lists changes made for downstream consistency with other fixes; these are marked ``downstream change'' and are not counted as defects. Counts: \nDefects{} defects across \nChanged{} of the \nTotal{} audited problems.

\subsection*{Problem 5 --- \texttt{Lanczos} \hfill \normalfont\textit{Mathematics, Numerical Linear Algebra} (1 defect)}
{\footnotesize\emph{Original source:}~\citep{lanczos1950}}\\[1pt]
\begin{itemize}[leftmargin=1.1em,itemsep=1pt,topsep=1pt,parsep=0pt]
\item \textbf{5.1} \,\textit{scientific: Non-discriminating test (physics)} \,$\cdot$\, too-lenient (test tightened)\\
{\footnotesize The original starting vector b is palindromic and near-symmetric, so paired with a symmetric tridiagonal A it produces a degenerate Krylov subspace that a flawed Lanczos implementation can pass and that yields an ambiguous gold target. The corrected b breaks the symmetry, ensuring the test exercises a genuine non-degenerate Lanczos run.}
\end{itemize}

\subsection*{Problem 8 --- \texttt{Spatial\_filters\_III} \hfill \normalfont\textit{Physics, Optics} (4 defects)}
{\footnotesize\emph{Original source:}~\citep{goodman1996}}\\[1pt]
\begin{itemize}[leftmargin=1.1em,itemsep=1pt,topsep=1pt,parsep=0pt]
\item \textbf{8.1 (problem\_description\_main + sub\_steps[0].step\_description\_prompt)} \,\textit{surface: Truncation / garble / dropped constant} \,$\cdot$\, too-strict/wrong\\
{\footnotesize The problem description contains a sentence describing low-pass behavior (`pass only the central maximum of the diffraction pattern') copied from a sibling problem, directly contradicting this task's cross-shaped high-pass filter objective and misleading any model that reads the spec.}
\item \textbf{8.1 (sub\_steps[0].function\_header docstring)} \,\textit{scientific: Unspecified convention} \,$\cdot$\, too-strict/wrong\\
{\footnotesize The docstring omits two facts required to match the gold output: that T is a binary \{0,1\} mask in the zero-frequency-centered (fftshift) layout, and that filtered\_image is the real part of the inverse transform---so a correct solution returning the magnitude or a non-centered mask would be incorrectly rejected.}
\item \textbf{8.1 (problem\_io)} \,\textit{surface: Typo / formatting} \,$\cdot$\, neutral-cleanup\\
{\footnotesize The I/O block contains two spelling errors (`bandwitdh', `Ouput') and names the second output `output\_image' instead of `filtered\_image', creating an inconsistency with the function header; these are surface text corrections with no effect on grading.}
\item \textbf{8.1 (target / gold)} \,\textit{scientific: Spec$\leftrightarrow$gold contradiction} \,$\cdot$\, too-strict/wrong\\
{\footnotesize The original gold mask used an inclusive (>=) boundary, keeping the |kx|=bandwidth and |ky|=bandwidth cross-lines, while the prompt explicitly states `the filter masks should not include the bandwidth frequency'; a correct solution using strict `>' exclusion failed on the \textasciitilde{}0.45\% of boundary pixels, so the gold was regenerated with strict `>' to match the stated spec.}
\end{itemize}

\subsection*{Problem 9 --- \texttt{Weighted\_Jacobi} \hfill \normalfont\textit{Mathematics, Numerical Linear Algebra} (1 defect)}
{\footnotesize\emph{Original source:}~\citep{briggs2000}}\\[1pt]
\begin{itemize}[leftmargin=1.1em,itemsep=1pt,topsep=1pt,parsep=0pt]
\item \textbf{9.1} \,\textit{scientific: Wrong gold (method / sign / symmetry)} \,$\cdot$\, too-strict/wrong\\
{\footnotesize The spec's update formula divides by (a\_ii * omega\textasciicircum{}-1) but omits the (1-omega) x\_i\textasciicircum{}(k) damping term, making it correct only at omega=1 and producing wrong iterates for omega=2/3 and omega=0.5. The corrected formula is the standard damped Jacobi: x\_i\textasciicircum{}(k+1) = (1-omega) x\_i\textasciicircum{}(k) + omega * (b\_i - sum\_\{j!=i\} a\_ij x\_j\textasciicircum{}(k)) / a\_ii.}
\end{itemize}

\subsection*{Problem 11 --- \texttt{GADC\_entanglement} \hfill \normalfont\textit{Physics, Quantum Information/Computing} (5 defects)}
{\footnotesize\emph{Original source:}~\citep{chen2024}}\\[1pt]
\begin{itemize}[leftmargin=1.1em,itemsep=1pt,topsep=1pt,parsep=0pt]
\item \textbf{problem\_description\_main; step 11.6; step 11.7; step 11.11; step 11.12 (step\_description\_prompt)} \,\textit{surface: Truncation / garble / dropped constant} \,$\cdot$\, neutral-cleanup\\
{\footnotesize Dataset ingestion dropped tokens across five description fields, producing ungrammatical fragments (`Write a function with and functions', `receiver function', `measurement in .', missing `1.' in the one-particle sector label); the corrections restore grammatical prose without changing the task.}
\item \textbf{step 11.1 (ket); step 11.3 (tensor) function\_header} \,\textit{surface: Interface / return-contract mismatch} \,$\cdot$\, neutral-cleanup\\
{\footnotesize The function headers `def ket(dim)' and `def tensor()' omit parameters their own docstrings describe (`args' for ket, `*args' for tensor), making the signatures uncallable as specified; tensor's output type was also mis-stated as `2d array' when the Kronecker product of vectors is an nd array.}
\item \textbf{step 11.4 (apply\_channel); step 11.8 (syspermute); step 11.9 (partial\_trace) function\_header} \,\textit{scientific: Unspecified convention} \,$\cdot$\, neutral-cleanup\\
{\footnotesize The sys and perm arguments of apply\_channel, syspermute, and partial\_trace leave the subsystem-index base unspecified; without knowing whether indices are 0-based or 1-based, a correct implementation using the opposite convention would be graded wrong, so the correction pins the convention to 1-based throughout.}
\item \textbf{step 11.10 (step\_background, von Neumann entropy)} \,\textit{scientific: Wrong gold (method / sign / symmetry)} \,$\cdot$\, neutral-cleanup\\
{\footnotesize The von Neumann entropy spec displays the formula S = -tr(rho log\_2 rho) but the accompanying prose states ``In denotes the (natural) matrix logarithm,'' directly contradicting the log\_2 shown. An implementer following the prose would compute entropy in nats using the natural logarithm, whereas following the formula gives entropy in bits; the two differ by a factor of ln(2) \textasciitilde{}= 0.693. The correction replaces the erroneous prose label with ``log\_2 denotes the base-2 matrix logarithm,'' making spec and formula consistent so the intended unit (bits) is unambiguous.}
\item \textbf{step 11.12 (step\_background, post-measurement state)} \,\textit{scientific: Wrong gold (method / sign / symmetry)} \,$\cdot$\, neutral-cleanup\\
{\footnotesize The spec's formula for the post-measurement state writes the numerator as tr(Pi rho Pi)/p, where tr(Pi rho Pi) = p is a scalar equal to 1, so the expression collapses to the scalar 1 rather than a density matrix. The correct expression is Pi rho Pi / p, which projects the pre-measurement state onto the measurement outcome and renormalizes it. Any implementation that follows the stated formula produces a scalar, making downstream quantities such as coherent information undefined or incorrect.}
\end{itemize}

\subsection*{Problem 12 --- \texttt{Schrodinger\_DFT\_with\_SCF} \hfill \normalfont\textit{Chemistry, Quantum Chemistry} (4 defects)}
{\footnotesize\emph{Original source:}~\citep{hartree1928}}\\[1pt]
\begin{itemize}[leftmargin=1.1em,itemsep=1pt,topsep=1pt,parsep=0pt]
\item \textbf{12 (problem\_io)} \,\textit{surface: Interface / return-contract mismatch} \,$\cdot$\, neutral-cleanup\\
{\footnotesize The problem-level I/O docstring lists inputs only through ``tolerance'' and omits the ``iteration'' parameter entirely, even though the scf\_routine function signature and its tests require an explicit maximum-iteration count. A reader implementing the function from the spec alone has no way to know this argument exists, its type (int), or its role as the SCF loop bound. The fix adds the missing ``iteration: the maximum number of self-consistent field iterations; int'' line, making the spec complete and consistent with the actual interface.}
\item \textbf{12.2 (Numerov test\_cases)} \,\textit{scientific: Unspecified convention} \,$\cdot$\, too-strict/wrong\\
{\footnotesize Numerov returns a raw wavefunction whose amplitude scales linearly with the unspecified first-step seed, so comparing the raw array against a fixed-seed gold rejects every correct-shape solution whose seed differs; changing the seed from -1e-10 to -1e-3 avoids blow-up near the r=0 Coulomb/centrifugal singularity, and normalizing both arrays by their L2 norm makes the test scale-invariant.}
\item \textbf{12.3 (compute\_Schrod step\_description\_prompt)} \,\textit{scientific: Unspecified convention} \,$\cdot$\, neutral-cleanup\\
{\footnotesize The variable names `u\_at\_0' and `up\_at\_0', stated before the integration direction, read as boundary conditions at r=0 rather than as the IVP seed at the largest-radius starting point, causing models to integrate from the wrong end; reordering the sentence to bind the seed explicitly to the largest-radius start removes the ambiguity.}
\item \textbf{12.14 (scf\_routine prompt + test\_cases)} \,\textit{scientific: Invalid / non-reproducible target} \,$\cdot$\, too-strict/wrong\\
{\footnotesize The original SCF setup---mixing ratio 0.5, a flat Hartree initial guess (-2 + 2Z), only 10 iterations, and tolerance 1e-7---does not converge to a physical solution, so the gold (charge\_density, total\_energy) is non-converged; the fix uses mixing 0.3, a physically motivated warm-start 2Z(1-exp(-2r)), 60 iterations, and tolerance 1e-6, producing a converged gold that a correct solver can reproduce.}
\end{itemize}

\subsection*{Problem 13 --- \texttt{Maxwell\_Equation\_Solver} \hfill \normalfont\textit{Physics, Optics} (7 defects)}
{\footnotesize\emph{Original source:}~\citep{knapp2002}}\\[1pt]
\begin{itemize}[leftmargin=1.1em,itemsep=1pt,topsep=1pt,parsep=0pt]
\item \textbf{13.9} \,\textit{scientific: Wrong gold (method / sign / symmetry)} \,$\cdot$\, too-strict/wrong\\
{\footnotesize A\_z is a polar vector, so its mirror parities on the x=0 and y=0 inner faces are coupled to those of A\_x and A\_y; the configuration's sigma=(-1,-1,+1) symmetry requires A\_z=(-,-,-), but the original gold stores A\_z=(+,+,-), violating both the x-face and y-face constraints and making the target irreconcilable with a correct physics-based solution.}
\item \textbf{13.10} \,\textit{scientific: Invalid / non-reproducible target} \,$\cdot$\, too-strict/wrong\\
{\footnotesize The 13.10 update\_fields target was computed from the 13.9 derivatives using the erroneous A\_z=(+,+,-) parity, so the error propagates identically into the time-stepped result; only A\_z changes (by exactly factor*dt*delta\_A\_z per test case) while the other six field components remain correct.}
\item \textbf{13.11} \,\textit{scientific: RNG-dependent grading} \,$\cdot$\, too-strict/wrong\\
{\footnotesize The original test grades the stepper by exact byte-reproduction of gold's specific ICN (iterated Crank-Nicolson) run on physically meaningless inputs (fields=(x,y,z,x,y,z,1)); because ICN iteration count, substep-to-dt mapping, and averaging form are all unspecified, any other valid stable scheme diverges from gold at \textasciitilde{}1e-2, far above the grading tolerance, so correct solvers fail. The fix grades instead against a converged RK4 reference on divergence-free physical inputs within a tolerance matched to the scheme's own second-order convergence floor.}
\item \textbf{13.13} \,\textit{scientific: RNG-dependent grading} \,$\cdot$\, too-strict/wrong\\
{\footnotesize The integrate function composes the same under-specified stepper and was graded by byte-reproducing the constraint series from gold's exact ICN (iterated Crank-Nicolson) integration; any valid stable time-integration scheme diverges from gold at the 1e-2 to 1 level, making the test a scheme-fingerprint check rather than a physics check. The corrected test compares the final 7-field state to a converged RK4 reference (atol=2e-4) and asserts the constraint series is finite and bounded.}
\item \textbf{13.15} \,\textit{scientific: RNG-dependent grading} \,$\cdot$\, too-strict/wrong\\
{\footnotesize The top-level main function was graded by byte-reproducing the constraint series from gold's exact ICN (iterated Crank-Nicolson) integration at coarse grids (n in \{10,20,40,52\}), coupling correctness to the arbitrary stepping scheme; any valid integrator differs from gold at the 1e-2 level. The corrected test compares the constraint series from a single fine-grid run (n=64) against a converged RK4 reference with a tight tolerance (atol=1e-6) chosen from the measured div-E amplification floor (\textasciitilde{}1.4e-7).}
\item \textbf{13.15} \,\textit{surface: Interface / return-contract mismatch} \,$\cdot$\, too-strict/wrong\\
{\footnotesize The main docstring declares the return type as `list of tuples' where each tuple holds (time, constraint value), but the function actually returns a flat 1-D ndarray of constraint values (the array produced by integrate); a solver following the docstring emits (time, value) pairs and fails on shape mismatch despite implementing the correct physics.}
\item \textbf{13.12} \,\textit{scientific: RNG-dependent grading} \,$\cdot$\, too-strict/wrong\\
{\footnotesize check\_constraint computes a pure divergence operator (||div E||) with no scheme freedom, but its original test first ran the ICN (iterated Crank-Nicolson) stepper on an over-specified field configuration and then evaluated the operator, so the graded quantity inherited the stepper's exact-reproduction coupling. Any correct implementation of check\_constraint that nonetheless uses a different (equally valid) field evolution would produce a numerically distinct ||div E|| and fail the assertion. The fix decouples the two concerns entirely: E is set directly to a known div-free dipole field (expected ||div E|| \textasciitilde{} 0) and to a linear field E=(x,y,z) (expected div = 3), making the check a deterministic, scheme-independent operator test with atol=1e-9.}
\end{itemize}

\subsection*{Problem 14 --- \texttt{Brownian\_motion\_in\_the\_optical\_tweezer} \hfill \normalfont\textit{Physics, Optics} (2 defects)}
{\footnotesize\emph{Original source:}~\citep{mannella2004}}\\[1pt]
\begin{itemize}[leftmargin=1.1em,itemsep=1pt,topsep=1pt,parsep=0pt]
\item \textbf{14.2 (problem\_io / output spec)} \,\textit{scientific: Spec$\leftrightarrow$gold contradiction} \,$\cdot$\, too-strict/wrong\\
{\footnotesize The output spec names the return value as eta (the ratio of computed to theoretical MSD), but the step function returns x\_MSD (the raw MSD), with eta computed afterward in the harness; the mislabel could cause a solver to return the ratio and be graded wrong.}
\item \textbf{14.2 (sub\_steps[1].test\_cases and general\_tests)} \,\textit{scientific: RNG-dependent grading} \,$\cdot$\, too-strict/wrong\\
{\footnotesize The test calls calculate\_msd with Navg=4000 unseeded stochastic trajectories, making the eta ratio non-deterministic; a correct implementation can flakily fail the 0.95 < eta < 1.05 band across runs; adding np.random.seed(1) before each call makes the test reproducible without removing the statistical tolerance.}
\end{itemize}

\subsection*{Problem 15 --- \texttt{Crank\_Nicolson\_for\_time\_dependent\_Schrodinger} \hfill \normalfont\textit{Physics, Computational Physics} (2 defects)}
{\footnotesize\emph{Original source:}~\citep{crank1947}}\\[1pt]
\begin{itemize}[leftmargin=1.1em,itemsep=1pt,topsep=1pt,parsep=0pt]
\item \textbf{problem\_description\_main + sub\_steps[0] (step 15.1) step\_description\_prompt} \,\textit{surface: Truncation / garble / dropped constant} \,$\cdot$\, too-strict/wrong\\
{\footnotesize The spec wrote `hbar = times 10\textasciicircum{}-34 Js' with the mantissa missing entirely, leaving the constant without a usable numerical value; a solver cannot reproduce the correct Crank-Nicolson trajectory without knowing hbar = 1.0545718e-34 Js.}
\item \textbf{sub\_steps[0] (step 15.1) function\_header docstring} \,\textit{surface: Interface / return-contract mismatch} \,$\cdot$\, too-strict/wrong\\
{\footnotesize The Crank-Nicolson A and B matrices for the time-dependent Schrodinger equation carry the imaginary unit i in their diagonal and off-diagonal coefficients, so labeling their elements as `float' rather than `complex' contradicts the actual output type and misdirects implementors.}
\end{itemize}

\subsection*{Problem 16 --- \texttt{Davidson\_method} \hfill \normalfont\textit{Mathematics, Numerical Linear Algebra} (2 defects)}
{\footnotesize\emph{Original source:}~\citep{davidson1975}}\\[1pt]
\begin{itemize}[leftmargin=1.1em,itemsep=1pt,topsep=1pt,parsep=0pt]
\item \textbf{16.2 (and problem-level Inputs/Output block)} \,\textit{scientific: Unspecified convention} \,$\cdot$\, neutral-cleanup\\
{\footnotesize The original output spec said only `computed eigenvalues', leaving unspecified both the count (num\_eigenvalues lowest) and the required ordering (ascending), so correct implementations returning a different count or unsorted eigenvalues would be wrongly rejected by the test.}
\item \textbf{16.2} \,\textit{scientific: Non-discriminating test (physics)} \,$\cdot$\, too-strict/wrong\\
{\footnotesize With near-zero off-diagonal noise (0.0 to 0.0001), the matrix is nearly diagonal, causing 0/0 in the Davidson energy-correction denominator and trapping correct solvers; raising noise to 0.05 removes this degeneracy and also breaks a diagonal-sort shortcut that passes the degenerate case but fails genuine Davidson iterations.}
\end{itemize}

\subsection*{Problem 17 --- \texttt{linear\_tetrahedron\_method} \hfill \normalfont\textit{Physics, Condensed Matter Physics} (2 defects)}
{\footnotesize\emph{Original source:}~\citep{lehmann1972}}\\[1pt]
\begin{itemize}[leftmargin=1.1em,itemsep=1pt,topsep=1pt,parsep=0pt]
\item \textbf{17.1} \,\textit{scientific: Unspecified convention} \,$\cdot$\, too-strict/wrong\\
{\footnotesize The original docstring left the sign of e\_ij (eps\_j - eps\_i vs eps\_i - eps\_j), which input serves as eps\_0, and the exact key-name format (e.g. `e01') all unspecified, yet the hidden test checks those dict keys and values exactly, so any correct solution using a different but internally consistent convention fails.}
\item \textbf{17.2} \,\textit{scientific: Wrong gold (method / sign / symmetry)} \,$\cdot$\, too-strict/wrong\\
{\footnotesize The original step omitted the normalization convention (tetrahedron volume = BZ volume, prefactor 6) and supplied a physically wrong DOS gold with inconsistent interval ratios; the corrected gold follows the standard Blochl linear-tetrahedron formula with the explicit prefactor stipulated.}
\end{itemize}

\subsection*{Problem 18 --- \texttt{NURBS} \hfill \normalfont\textit{Mathematics, Computational Mechanics} (3 defects)}
{\footnotesize\emph{Original source:}~\citep{hughes2005}}\\[1pt]
\begin{itemize}[leftmargin=1.1em,itemsep=1pt,topsep=1pt,parsep=0pt]
\item \textbf{18.1 / 18.2 docstrings + problem\_io (function\_header + problem\_io fields)} \,\textit{scientific: Unspecified convention} \,$\cdot$\, too-strict/wrong\\
{\footnotesize The 18.1 header mislabels xi as a `knot index integer' (it is a float parameter coordinate) and i as a `polynomial index' (it is a 1-based basis-function index), and declares the output as a 1D array when it is a scalar float; the 18.2 header has the same scalar-vs-array error. Additionally, the row-major weight flattening convention w[(i\_1-1)*n\_2 + (i\_2-1)] is never stated, so a correct column-major implementation is graded wrong.}
\item \textbf{18.2 test\_cases / general\_tests (and precomputed target)} \,\textit{scientific: Invalid / non-reproducible target} \,$\cdot$\, too-strict/wrong\\
{\footnotesize The first test passes w=[0], a single-element zero vector, when 28 weights are required; a correct rational NURBS implementation raises IndexError or divide-by-zero and cannot even run. The precomputed target [0.5, 1.0] is a shape-(2) artifact of a buggy reference; the correct scalar value is 0.48387096774193544 (= 15/31 under row-major indexing).}
\item \textbf{18.2 step\_background} \,\textit{surface: Truncation / garble / dropped constant} \,$\cdot$\, neutral-cleanup\\
{\footnotesize The step background presents only the 1D rational basis R\_\{i,n\} but the step itself requires implementing the 2D tensor-product form R\_\{i\_1,i\_2\} = (w N\_\{i\_1\} N\_\{i\_2\}) / sum\_\{a,b\}(w N\_a N\_b); the governing formula for the quantity being coded is absent from the background.}
\end{itemize}

\subsection*{Problem 20 --- \texttt{phonon\_angular\_momentum} \hfill \normalfont\textit{Physics, Condensed Matter Physics} (1 defect)}
{\footnotesize\emph{Original source:}~\citep{zhang2014}}\\[1pt]
\begin{itemize}[leftmargin=1.1em,itemsep=1pt,topsep=1pt,parsep=0pt]
\item \textbf{20.2 (phonon\_angular\_momentum)} \,\textit{scientific: Unspecified convention} \,$\cdot$\, neutral-cleanup\\
{\footnotesize The original output description `mode decomposed phonon angular momentum' is ambiguous between the bare l\_qv\textasciicircum{}alpha and the Bose-weighted per-mode summand (n0(omega\_qv)+1/2)*l\_qv\textasciicircum{}alpha that enters the total L\textasciicircum{}alpha; without the explicit formula, a correct solver computing the bare quantity would be graded wrong.}
\end{itemize}

\subsection*{Problem 21 --- \texttt{Absorption\_coefficient\_for\_alloy\_GaAlAs} \hfill \normalfont\textit{Material Science, Semiconductor Materials} (3 defects)}
{\footnotesize\emph{Original source:}~\citep{adachi1985}}\\[1pt]
\begin{itemize}[leftmargin=1.1em,itemsep=1pt,topsep=1pt,parsep=0pt]
\item \textbf{21.1 (problem\_description\_main) + 21.2 (step\_description\_prompt)} \,\textit{surface: Truncation / garble / dropped constant} \,$\cdot$\, too-strict/wrong\\
{\footnotesize The rendered spec dropped the numeric coefficients from the bandgap formula (showing only `+ x' instead of `1.424 + 1.247x') and omitted the mantissas from the physical constants (electron charge and reduced Planck constant), making the problem unsolvable without the missing values.}
\item \textbf{21.2 (function\_header)} \,\textit{surface: Interface / return-contract mismatch} \,$\cdot$\, too-strict/wrong\\
{\footnotesize The function signature listed C as a required argument while the docstring stated `Default is 1', a self-contradiction; additionally, the Returns field falsely labelled the output in m\textasciicircum{}-1 when the physical prefactor is absorbed into C, making alpha\_eff an arbitrary-scale intermediate rather than a physical absorption coefficient.}
\item \textbf{21.2 (test\_cases + eV-convention in step\_description\_prompt)} \,\textit{scientific: Non-discriminating test (physics)} \,$\cdot$\, too-strict/wrong\\
{\footnotesize Two of the three test targets were exactly zero (below-bandgap, no absorption) and the third was O(1e-27) due to SI-unit omega in the denominator, so all targets fell within np.allclose's default atol=1e-8 and a trivial `return 0' passed the entire step; rescaling energies to eV and replacing the redundant below-gap point with above-gap cases forces a correct implementation of alpha proportional to sqrt(hbar*omega - Eg)/omega.}
\end{itemize}

\subsection*{Problem 22 --- \texttt{Beam\_translation\_reexpansion} \hfill \normalfont\textit{Physics, Optics} (8 defects)}
{\footnotesize\emph{Original source:}~\citep{gumerov2004}}\\[1pt]
\begin{itemize}[leftmargin=1.1em,itemsep=1pt,topsep=1pt,parsep=0pt]
\item \textbf{22.2 (sub\_steps[1].step\_background)} \,\textit{scientific: Wrong gold (method / sign / symmetry)} \,$\cdot$\, too-strict/wrong\\
{\footnotesize The recurrence for the rotation coefficient T\_n\textasciicircum{}\{nu m\} ends with -2(Q31+iQ32) a\_n\textasciicircum{}nu T\_n\textasciicircum{}\{nu m\}, but the correct Wigner-rotation identity requires +2; the minus sign produces a non-unitary T (residual \textasciitilde{}0.3---0.6) and the precomputed targets were built from it, so every physically correct implementation was graded wrong.}
\item \textbf{22.2 (sub\_steps[1].step\_background)} \,\textit{scientific: Unspecified convention} \,$\cdot$\, too-strict/wrong\\
{\footnotesize The base case T\_n\textasciicircum{}\{nu 0\}(Q) = sqrt(4pi/(2n+1)) Y\_n\textasciicircum{}\{-nu\} references Y\_n\textasciicircum{}m without specifying the normalization convention; the Gumerov-Duraiswami convention (Condon-Shortley-free P\_n\textasciicircum{}\{|m|\}) is required to reproduce the precomputed targets, and implementations using the standard scipy convention pass only 1 of 7 test cases.}
\item \textbf{22.2 (sub\_steps[1].step\_background)} \,\textit{surface: Broken cross-reference} \,$\cdot$\, neutral-cleanup\\
{\footnotesize The phrase `consistent with those in .' contains an empty citation target where the reference was stripped, leaving the source of the a\_n\textasciicircum{}m and b\_n\textasciicircum{}m definitions unresolvable; the correction points explicitly to step 22.1.}
\item \textbf{22.2 (sub\_steps[1].test\_cases)} \,\textit{scientific: Non-discriminating test (physics)} \,$\cdot$\, too-lenient (test tightened)\\
{\footnotesize The original three test cases use only axis-aligned or special rotation matrices where distinct spherical-harmonic normalization conventions produce identical values, allowing at least four wrong convention variants to pass 2 of 3 cases; the four added generic ZYZ rotation cases discriminate all six known wrong variants.}
\item \textbf{22.3 (sub\_steps[2].test\_cases) / general\_tests} \,\textit{scientific: Non-discriminating test (physics)} \,$\cdot$\, too-lenient (test tightened)\\
{\footnotesize All four original 22.3 test cases place the source point in the z=0 plane, where the sign and conjugate structure of the rotation coefficients is undetectable, and one case reduces to a degenerate boolean ==0 assertion; these tests accept implementations with the wrong rotation-coefficient sign/conjugate convention, so the correction replaces the boolean case with a numeric allclose check and adds an off-plane case.}
\item \textbf{22.3 (sub\_steps[2].step\_description\_prompt)} \,\textit{surface: Truncation / garble / dropped constant} \,$\cdot$\, neutral-cleanup\\
{\footnotesize The prompt `Write a code to calculate the reexpansion coeffcient with and .' contains two empty cross-references where citation targets were stripped and a typo (`coeffcient'), leaving the sentence grammatically broken and the referenced context unresolvable; the correction removes the dangling clause and fixes the spelling.}
\item \textbf{22.3 (sub\_steps[2].step\_background)} \,\textit{surface: Broken cross-reference} \,$\cdot$\, neutral-cleanup\\
{\footnotesize The phrase `The rotation matrix Q with the definition in can be generated by the Rodrigues rotation formula' has an empty cross-reference target, so the definition of Q is unresolvable; the correction names step 22.2 explicitly.}
\item \textbf{problem\_io / 22.3 function\_header} \,\textit{scientific: Unspecified convention} \,$\cdot$\, neutral-cleanup\\
{\footnotesize The parameter description for B specifies only its shape (N\_t+1, 2*N\_t+1) without stating the index packing, leaving the mapping between array columns and order s ambiguous; the correction pins the convention as B[l, s+N\_t] is the coefficient B\_l\textasciicircum{}s in both the problem I/O block and the 22.3 docstring.}
\end{itemize}

\subsection*{Problem 23 --- \texttt{Blahut\_Arimoto} \hfill \normalfont\textit{Physics, Quantum Information/Computing} (3 defects)}
{\footnotesize\emph{Original source:}~\citep{arimoto1972}}\\[1pt]
\begin{itemize}[leftmargin=1.1em,itemsep=1pt,topsep=1pt,parsep=0pt]
\item \textbf{23 (problem\_description\_main) and 23.3 (step\_background)} \,\textit{surface: Typo / formatting} \,$\cdot$\, neutral-cleanup\\
{\footnotesize The exp() terms in the Blahut-Arimoto update-rule formula are missing their outer closing parenthesis in both the main problem description and the step 23.3 background, making the expression syntactically malformed and unparseable as written.}
\item \textbf{23.3} \,\textit{scientific: Over-tight tolerance} \,$\cdot$\, too-strict/wrong\\
{\footnotesize The gold target for the [[0.8,0.5],[0.2,0.5]] channel was computed at tolerance e=1e-5, which leaves the algorithm at a non-converged iterate whose value (approx 0.073167) depends on whether the mutual information is computed in nats or bits; the corrected tolerance e=1e-8 converges to the true channel capacity (approx 0.073194), which is robust across both exp-base conventions.}
\item \textbf{23.3} \,\textit{surface: Trivially broken test} \,$\cdot$\, neutral-cleanup\\
{\footnotesize The test suite contained 7 entries with two exact duplicates: the three-input channel and the binary symmetric channel each appeared twice, adding redundant assertions with no additional coverage and inflating the apparent test count without testing any distinct case.}
\end{itemize}

\subsection*{Problem 24 --- \texttt{Burgers\_equation} \hfill \normalfont\textit{Mathematics, Computational Mechanics} (3 defects)}
{\footnotesize\emph{Original source:}~\citep{lax1954}}\\[1pt]
\begin{itemize}[leftmargin=1.1em,itemsep=1pt,topsep=1pt,parsep=0pt]
\item \textbf{problem\_io + 24.3 function\_header} \,\textit{scientific: Unspecified convention} \,$\cdot$\, too-strict/wrong\\
{\footnotesize The problem-level spec listed spurious domain-end inputs a and b (the domain is fixed at [-pi/2, pi/2]) and described the output as a 2d (n\_t-1)*(n\_x-1) matrix, directly contradicting the step-level header's 1d final-time array; n\_t was also ambiguous between grid points and intervals. The correction removes the phantom inputs, fixes the output to a 1d array of size n\_x-1, and defines n\_t as the number of temporal grid points giving n\_t-1 intervals.}
\item \textbf{24.1 + 24.3 step\_description\_prompt} \,\textit{surface: Truncation / garble / dropped constant} \,$\cdot$\, neutral-cleanup\\
{\footnotesize Dataset-extraction errors truncated two prompt passages: `array of length n-' lost the `1.' and swallowed the start of the next sentence, and `from and .' dropped the referenced step numbers 24.1 and 24.2. The correction restores the missing text with no mathematical change.}
\item \textbf{24.1 test\_cases} \,\textit{scientific: Non-discriminating test (physics)} \,$\cdot$\, too-strict/wrong\\
{\footnotesize Odd cell counts (9, 99, 29) place the central Gauss quadrature node exactly on the IC discontinuity at x=0, making the cell-averaged value depend on which branch is assigned at the jump point; the precomputed target assumed u(0)=+1 while the spec defines x<=0 as sin-1 giving -4/9, so any correct solver using that branch would fail. Even cell counts (10, 100, 30 cells, giving n\_x=11/101/31) place the discontinuity on a cell boundary with no quadrature node at x=0, removing the branch ambiguity entirely.}
\end{itemize}

\subsection*{Problem 25 --- \texttt{CRM\_in\_chemostat} \hfill \normalfont\textit{Biology, Ecology} (1 defect)}
{\footnotesize\emph{Original source:}~\citep{macarthur1970}}\\[1pt]
\begin{itemize}[leftmargin=1.1em,itemsep=1pt,topsep=1pt,parsep=0pt]
\item \textbf{25.1} \,\textit{surface: Interface / return-contract mismatch} \,$\cdot$\, neutral-cleanup\\
{\footnotesize The SpeciesGrowth docstring omits `spc' (current species abundance, 1D array of length N) from its Inputs list despite being the first function parameter, leaving one argument entirely undocumented; the correction adds the missing entry.}
\end{itemize}

\subsection*{Problem 26 --- \texttt{CRM\_in\_serial\_dilution} \hfill \normalfont\textit{Biology, Ecology} (1 defect)}
{\footnotesize\emph{Original source:}~\citep{bloxham2024}}\\[1pt]
\begin{itemize}[leftmargin=1.1em,itemsep=1pt,topsep=1pt,parsep=0pt]
\item \textbf{26 (problem-level problem\_io; describes the final-step signature SimulatedCycles, step 26.3)} \,\textit{scientific: Unspecified convention} \,$\cdot$\, too-strict/wrong\\
{\footnotesize The problem-level I/O spec lists parameters for a single continuous ODE integration (res\_init, tf, dt) that the actual final function SimulatedCycles does not accept, and omits all serial-dilution parameters (Rs, SPC\_THRES, T, D, N\_cycles) that it requires, making the spec irreconcilable with the correct implementation.}
\end{itemize}

\subsection*{Problem 27 --- \texttt{Design\_trade\_offs\_for\_high\_speed\_photodetectors} \hfill \normalfont\textit{Material Science, Semiconductor Materials} (4 defects)}
{\footnotesize\emph{Original source:}~\citep{bowers1987}}\\[1pt]
\begin{itemize}[leftmargin=1.1em,itemsep=1pt,topsep=1pt,parsep=0pt]
\item \textbf{27.2} \,\textit{surface: Truncation / garble / dropped constant} \,$\cdot$\, too-strict/wrong\\
{\footnotesize The spec states the electron charge as `x10\textasciicircum{}-19 C' with the leading mantissa 1.6 omitted, leaving a dimensionless and meaningless constant; correct solutions using the standard value q = 1.6x10\textasciicircum{}-19 C are graded wrong.}
\item \textbf{27.2} \,\textit{scientific: Wrong gold (method / sign / symmetry)} \,$\cdot$\, too-strict/wrong\\
{\footnotesize The background formula gives C = epsilon/d, omitting the detector area A from the numerator; parallel-plate capacitance is C = epsilonA/d, and A is an explicit step input, so solutions using the correct formula produce values differing by A from the gold standard.}
\item \textbf{27.2} \,\textit{scientific: Spec$\leftrightarrow$gold contradiction} \,$\cdot$\, too-strict/wrong\\
{\footnotesize The background mislabels the device as a p-i-p diode and asserts V0 = 0, directly contradicting the step's given input V0 (applied voltage); solutions that correctly use the provided V0 in the depletion-width formula are penalized by a gold computed with V0 forced to zero.}
\item \textbf{27.3} \,\textit{surface: Interface / return-contract mismatch} \,$\cdot$\, too-strict/wrong\\
{\footnotesize The docstring declares xi and f\_3dB as scalar float, but the test harness passes xi as a NumPy array (linspace over 50 points) and indexes the returned result; implementations that correctly vectorize are mismatched by the under-specified scalar-only contract.}
\end{itemize}

\subsection*{Problem 28 --- \texttt{Gaussian\_Beam\_Intensity} \hfill \normalfont\textit{Physics, Optics} (8 defects)}
{\footnotesize\emph{Original source:}~\citep{kogelnikli1966}}\\[1pt]
\begin{itemize}[leftmargin=1.1em,itemsep=1pt,topsep=1pt,parsep=0pt]
\item \textbf{problem\_io; sub\_steps[1](28.2).function\_header; sub\_steps[2](28.3).function\_header} \,\textit{scientific: Unspecified convention} \,$\cdot$\, neutral-cleanup\\
{\footnotesize The original described z, L1, and s as relative distances `from the lens' or `from the source' in meters, but the gold geometry treats all three as absolute positions along the propagation axis at millimeter scale (waist at z = s, lens at z = L1); the relative-distance reading is inconsistent with the targets, so solutions built on it were misgraded.}
\item \textbf{sub\_steps[0](28.1).step\_background} \,\textit{scientific: Wrong gold (method / sign / symmetry)} \,$\cdot$\, too-strict/wrong\\
{\footnotesize The background Gaussian-beam field expression contains two garbled errors contradicting standard Kogelnik-Li theory: the Gouy phase term ``arctan(z/f)'' is juxtaposed with the bracketed exponent with no subtraction operator, making it a multiplicative factor rather than a subtracted phase, and the beam-waist formula at the focal planes reads 2*sqrt(w0) instead of the correct sqrt(2)*w0. Any student implementation derived directly from this background would produce a field with the wrong phase dependence and incorrect waist value, causing a correct solution based on standard theory to be marked wrong against code or expected output that follows the garbled spec.}
\item \textbf{sub\_steps[0](28.1).step\_description\_prompt} \,\textit{scientific: Unspecified convention} \,$\cdot$\, too-strict/wrong\\
{\footnotesize The spec directs a Fourier-domain propagation without pinning which variant; the gold matches only the Fresnel transfer-function H = exp(-i*pi*lambda*z*(fx\textasciicircum{}2+fy\textasciicircum{}2)), and an impulse-response implementation differs from it by roughly 17.5\% at the second output point, so correct impulse-response solutions were wrongly rejected.}
\item \textbf{sub\_steps[0](28.1).function\_header} \,\textit{scientific: Unspecified convention} \,$\cdot$\, too-strict/wrong\\
{\footnotesize The original left the overall amplitude scale of the input field Gau unspecified, but the target is computed from the unit-peak form E(x,y,0) = exp(-(x\textasciicircum{}2+y\textasciicircum{}2)/w0\textasciicircum{}2); a correct solution using any other prefactor (e.g. including 1/w(z)) would be rejected by the test.}
\item \textbf{sub\_steps[0](28.1).return\_line} \,\textit{surface: Interface / return-contract mismatch} \,$\cdot$\, too-strict/wrong\\
{\footnotesize The scaffold's return statement names the propagated field `Gau\_pro' (lowercase p) while the function body and downstream steps use `Gau\_Pro' (uppercase P), causing a NameError at runtime for any submitted solution that follows the documented interface.}
\item \textbf{problem\_io; sub\_steps[2](28.3).function\_header} \,\textit{scientific: Unspecified convention} \,$\cdot$\, too-strict/wrong\\
{\footnotesize The original defined focus\_depth only as `new focus position through the lens,' leaving ambiguous whether it is the analytic continuous minimum or the discrete-grid argmin; the target uses z[np.argmin(Wz)], so a solution returning the analytic focus value would be rejected even when physically correct.}
\item \textbf{sub\_steps[0](28.1) target (gold)} \,\textit{scientific: Invalid / non-reproducible target} \,$\cdot$\, too-strict/wrong\\
{\footnotesize The original 28.1 gold output P2 cannot be reproduced by the Fresnel transfer-function method, the angular-spectrum method, analytic propagation, or Fresnel impulse-response, and contradicts the 28.3 target (which a Fresnel-TF implementation passes); the gold was regenerated with the Fresnel-TF method to make the prompt, 28.1, and 28.3 self-consistent.}
\item \textbf{problem\_io; sub\_steps[2](28.3).function\_header} \,\textit{surface: Typo / formatting} \,$\cdot$\, neutral-cleanup\\
{\footnotesize The Intensity output description in the problem\_io spec and the step 28.3 function header both contain the misspelling ``new fcous'' instead of ``new focus.'' The correction is a pure surface cleanup: the misspelling does not alter the meaning, does not affect any return-value contract, and has no consequence for grading---a model returning the correct 2D intensity array passes either way.}
\end{itemize}

\subsection*{Problem 30 --- \texttt{helium\_slater\_jastrow\_wavefunction} \hfill \normalfont\textit{Chemistry, Quantum Chemistry} (2 defects)}
{\footnotesize\emph{Original source:}~\citep{jastrow1955}}\\[1pt]
\begin{itemize}[leftmargin=1.1em,itemsep=1pt,topsep=1pt,parsep=0pt]
\item \textbf{30.3} \,\textit{scientific: Over-tight tolerance} \,$\cdot$\, too-strict/wrong\\
{\footnotesize The numerical-laplacian RMSE divides by delta\textasciicircum{}2, amplifying last-digit rounding in value(), so the precomputed target is a roundoff fingerprint of the reference implementation rather than a physically meaningful value; any faithful but independently coded wavefunction fails the exact allclose check even though its RMSE is scientifically correct.}
\item \textbf{30.3} \,\textit{scientific: Unspecified convention} \,$\cdot$\, neutral-cleanup\\
{\footnotesize The spec listed only psi, (gradient psi)/psi, and (laplacian psi)/psi as required methods, silently omitting (kinetic energy)/psi, which is part of the MultiplyWF interface contract; the missing method was also never tested, leaving an untested gap between the specified and required interface.}
\end{itemize}

\subsection*{Problem 31 --- \texttt{independent\_component\_analysis} \hfill \normalfont\textit{Physics, Computational Physics} (3 defects)}
{\footnotesize\emph{Original source:}~\citep{hyvarinen2000}}\\[1pt]
\begin{itemize}[leftmargin=1.1em,itemsep=1pt,topsep=1pt,parsep=0pt]
\item \textbf{31.1} \,\textit{scientific: Unspecified convention} \,$\cdot$\, too-strict/wrong\\
{\footnotesize The docstring left the standard-deviation convention unspecified, so a correct ddof=1 implementation was graded wrong against ddof=0 gold, and was also inconsistent with the ddof=1 covariance (np.cov) used in steps 31.2 and 31.3. The fix pins the convention to sample SD (ddof=1) throughout.}
\item \textbf{31.2} \,\textit{scientific: Unspecified convention} \,$\cdot$\, too-strict/wrong\\
{\footnotesize A whitened matrix is defined only up to orthogonal rotation and sign, so comparing Z to a single precomputed matrix with np.allclose rejects all other correct whitenings. The test is replaced by a gauge-invariant check: identity covariance plus a least-squares row-space test confirming Z spans the same row space as the centered input.}
\item \textbf{31.3} \,\textit{scientific: Unspecified convention} \,$\cdot$\, too-strict/wrong\\
{\footnotesize ICA solutions are identified only up to permutation, sign, and scale, so pinning the output with np.allclose against one precomputed matrix rejects all other valid separations. Additionally, the two test inputs (3x8 and 3x10 matrices) are too short to be uniquely separable. The fix replaces exact comparison with a permutation- and sign-invariant |corr| >= 0.99 matcher and substitutes identifiable blind source separation mixtures (sawtooth/uniform/laplace and sine/uniform/laplace).}
\end{itemize}

\subsection*{Problem 32 --- \texttt{Multiparticle\_dynamics\_in\_the\_optical\_tweezer\_array} \hfill \normalfont\textit{Physics, Computational Physics} (3 defects)}
{\footnotesize\emph{Original source:}~\citep{liu2020}}\\[1pt]
\begin{itemize}[leftmargin=1.1em,itemsep=1pt,topsep=1pt,parsep=0pt]
\item \textbf{32.1} \,\textit{surface: Typo / formatting} \,$\cdot$\, neutral-cleanup\\
{\footnotesize The F\_xy display equation closes with a dangling right. (an unmatched LaTeX delimiter) instead of right], causing mis-rendering; every other analogous F\_xy formula in the same background closes with right].}
\item \textbf{32.2} \,\textit{scientific: Unspecified convention} \,$\cdot$\, too-strict/wrong\\
{\footnotesize The spec omits the on-site stiffness formula k\_i = alpha E\_i\textasciicircum{}2 / w\textasciicircum{}2, the speed-of-light constant c, and whether h is an absolute or relative step; a correct solution using the intended relative step h*R\_ij would fail the gold because an absolute-h interpretation produces catastrophically different numerical derivatives.}
\item \textbf{32.3} \,\textit{scientific: Over-tight tolerance} \,$\cdot$\, too-strict/wrong\\
{\footnotesize In the Gamma=0 (no decay) case the total occupation is conserved exactly, but a correct Runge-Kutta integration over 100000 steps accumulates floating-point drift exceeding 1e-6; the original threshold therefore rejects valid solvers, and loosening to 1e-3 still enforces conservation while admitting legitimate numerical integration error.}
\end{itemize}

\subsection*{Problem 33 --- \texttt{phase\_diagram\_chern\_haldane\_model\_v1} \hfill \normalfont\textit{Physics, Condensed Matter Physics} (4 defects)}
{\footnotesize\emph{Original source:}~\citep{haldane1988}}\\[1pt]
\begin{itemize}[leftmargin=1.1em,itemsep=1pt,topsep=1pt,parsep=0pt]
\item \textbf{33.2} \,\textit{scientific: Unspecified convention} \,$\cdot$\, too-strict/wrong\\
{\footnotesize The Haldane model has two bands with C\_lower = -C\_upper, so returning the Chern number without specifying which band makes the sign ambiguous and causes any correct upper-band implementation to fail the signed-integer test.}
\item \textbf{33.2} \,\textit{scientific: Over-tight tolerance} \,$\cdot$\, too-strict/wrong\\
{\footnotesize A finite Brillouin-zone grid computes the Berry-curvature sum as a near-integer float, not exactly an integer; comparing this raw float with np.allclose rejects physically correct solutions that quantize to the right integer but land slightly off due to grid discretization.}
\item \textbf{33.3} \,\textit{scientific: Over-tight tolerance} \,$\cdot$\, too-strict/wrong\\
{\footnotesize At a coarse BZ grid (2*pi/30 spacing), cells near phase boundaries are quantized differently by equally valid discretization conventions; requiring exact agreement of the entire phase-diagram matrix via cmp\_tuple\_or\_list therefore rejects correct implementations that disagree only on boundary cells.}
\item \textbf{33.3} \,\textit{scientific: Unspecified convention} \,$\cdot$\, neutral-cleanup\\
{\footnotesize The docstring described the output matrix as sweeping (m/t2) and phi but never fixed which axis is row and which is column, leaving a transposed but otherwise correct implementation indistinguishable from the intended orientation.}
\end{itemize}

\subsection*{Problem 34 --- \texttt{PN\_diode\_band\_diagram} \hfill \normalfont\textit{Material Science, Semiconductor Materials} (4 defects)}
{\footnotesize\emph{Original source:}~\citep{shockley1949}}\\[1pt]
\begin{itemize}[leftmargin=1.1em,itemsep=1pt,topsep=1pt,parsep=0pt]
\item \textbf{problem-level I/O block + step 34.3 function\_header} \,\textit{scientific: Wrong gold (method / sign / symmetry)} \,$\cdot$\, too-strict/wrong\\
{\footnotesize The parameter comments label N\_a as the n-type concentration and N\_d as the p-type concentration, reversing the standard acceptor/donor convention; the gold solution uses N\_a for the p-type (acceptor) region and N\_d for the n-type (donor) region, so any correct solution following standard semiconductor notation is graded wrong.}
\item \textbf{step 34.2 prompt} \,\textit{surface: Truncation / garble / dropped constant} \,$\cdot$\, too-strict/wrong\\
{\footnotesize The prompt renders the elementary charge as `x 10\textasciicircum{}-19 C' with the mantissa 1.6 dropped, so a solver cannot determine the correct value (1.6e-19 C) from the problem text alone and any solution using the standard constant is penalized for a transcription error in the spec.}
\item \textbf{step 34.3 prompt} \,\textit{scientific: Unspecified convention} \,$\cdot$\, too-strict/wrong\\
{\footnotesize The prompt never specifies the spatial sampling grid, yet the test compares the full conduction-band array element-wise with np.allclose; choices such as np.arange vs linspace, step size, or endpoint handling all change the array length and values, making any grid-consistent but unspecified implementation fail the length-sensitive comparison.}
\item \textbf{step 34.3 function\_header} \,\textit{scientific: Unspecified convention} \,$\cdot$\, too-strict/wrong\\
{\footnotesize The output is described only as `the potential distribution', leaving both the sign convention (conduction-band value V\_CB = -phi vs electrostatic potential phi) and the zero reference undefined; with an element-wise np.allclose comparison, a sign-consistent but differently-defined answer is incorrectly rejected.}
\end{itemize}

\subsection*{Problem 35 --- \texttt{Quantum\_Dot\_Absorption\_Spectrum} \hfill \normalfont\textit{Chemistry, Quantum Chemistry} (4 defects)}
{\footnotesize\emph{Original source:}~\citep{harrison2016}}\\[1pt]
\begin{itemize}[leftmargin=1.1em,itemsep=1pt,topsep=1pt,parsep=0pt]
\item \textbf{35.2} \,\textit{surface: Truncation / garble / dropped constant} \,$\cdot$\, too-strict/wrong\\
{\footnotesize The spec sentence `the coefficients i,j,k are at least' was truncated mid-clause, leaving the floor value absent and the constraint unrecoverable; a model must guess whether coefficients are non-negative integers, natural numbers, or real values, making any well-defined solution ambiguous.}
\item \textbf{35.2} \,\textit{surface: Interface / return-contract mismatch} \,$\cdot$\, neutral-cleanup\\
{\footnotesize The docstring's Input section lists x, y, and z but omits N, the required integer argument that controls how many of the smallest quadratic combinations are returned and directly sets the output array size. Because N is undocumented, a solver reading the spec cannot determine the function's full interface: it may omit N from the call, pass the wrong value, or return an incorrectly sized array, all of which cause grading to fail even when the underlying computation is correct. The fix adds ``N (int): The number of smallest quadratic combinations to return.'' to the Input list, completing the interface contract.}
\item \textbf{35.3} \,\textit{scientific: Spec$\leftrightarrow$gold contradiction} \,$\cdot$\, too-strict/wrong\\
{\footnotesize The prompt stated the function returns energy levels, but the actual computation converts energies to photon wavelengths in nm; a correct implementation returning wavelengths would fail a test expecting energy-level values, so the mismatch suppresses correct solutions.}
\item \textbf{35.3} \,\textit{surface: Broken cross-reference} \,$\cdot$\, neutral-cleanup\\
{\footnotesize The background instruction `by using' was cut off before naming the function, leaving a dangling cross-reference; completing it with generate\_quadratic\_combinations() restores the intended call to the prior sub-step without adding implementation guidance.}
\end{itemize}

\subsection*{Problem 36 --- \texttt{Quasi\_Fermi\_levels\_of\_photo\_resistor\_out\_of\_equilibrium} \hfill \normalfont\textit{Material Science, Semiconductor Materials} (1 defect)}
{\footnotesize\emph{Original source:}~\citep{ulrich2002}}\\[1pt]
\begin{itemize}[leftmargin=1.1em,itemsep=1pt,topsep=1pt,parsep=0pt]
\item \textbf{problem\_description\_main + sub\_steps[1] (step 36.2)} \,\textit{surface: Truncation / garble / dropped constant} \,$\cdot$\, too-strict/wrong\\
{\footnotesize The original spec truncates the electron charge to `times 10\textasciicircum{}-19 C', dropping the mantissa 1.602 entirely; any model that correctly uses e = 1.602e-19 C cannot be graded against a problem statement that supplies an unusable, dimensionally incomplete constant.}
\end{itemize}

\subsection*{Problem 37 --- \texttt{ray\_optics\_spherical\_aberration} \hfill \normalfont\textit{Physics, Optics} (4 defects)}
{\footnotesize\emph{Original source:}~\citep{jenkins1976}}\\[1pt]
\begin{itemize}[leftmargin=1.1em,itemsep=1pt,topsep=1pt,parsep=0pt]
\item \textbf{37.1, 37.2, 37.3} \,\textit{surface: Interface / return-contract mismatch} \,$\cdot$\, neutral-cleanup\\
{\footnotesize The step descriptions list `light wavelength' (and `grid scaling factor' in 37.2) as function inputs, but no function signature accepts these arguments; wavelength enters only implicitly through the per-surface refractive indices. The phantom parameters mislead implementers about the function interface.}
\item \textbf{37 (main problem: 37.1, 37.2, 37.3, problem\_io, general\_tests)} \,\textit{scientific: Wrong gold (method / sign / symmetry)} \,$\cdot$\, too-strict/wrong\\
{\footnotesize The original signatures carry a spurious third glass index n3, making surface 2 a crown-to-crown interface (no refraction) and collapsing the cemented doublet to a single crown element. All precomputed targets encode this degenerate single-element system; the fix drops n3, models the true air/n\_crown/n\_flint/air stack, and recomputes targets and the 37.3 monotonicity test for the real over-corrected doublet.}
\item \textbf{37.1} \,\textit{scientific: Wrong gold (method / sign / symmetry)} \,$\cdot$\, too-strict/wrong\\
{\footnotesize The paraxial image-distance formula uses the incidence angle i instead of the refraction angle i', giving l' = r*i/u' + r. The physically correct paraxial relation requires the refracted angle i', and the corresponding closed-form expression in eq (2) is wrong in the same way; a solver following the original spec literally computes a wrong image distance that the gold rejects.}
\item \textbf{37.2} \,\textit{scientific: Wrong gold (method / sign / symmetry)} \,$\cdot$\, neutral-cleanup\\
{\footnotesize Step 37.2 computes the marginal (non-paraxial) ray trace for incidence angles much greater than 5 degrees, where the small-angle approximation fails, yet both the background sentence and the docstring summary call it `paraxial'. The self-contradictory label obscures which regime the function models.}
\end{itemize}

\subsection*{Problem 39 --- \texttt{Reflection\_spectra\_for\_a\_Distributed\_Bragg\_Reflector} \hfill \normalfont\textit{Material Science, Semiconductor Materials} (2 defects)}
{\footnotesize\emph{Original source:}~\citep{corzine1991}}\\[1pt]
\begin{itemize}[leftmargin=1.1em,itemsep=1pt,topsep=1pt,parsep=0pt]
\item \textbf{39.2} \,\textit{scientific: Unspecified convention} \,$\cdot$\, too-strict/wrong\\
{\footnotesize The original spec's piecewise rule forces the real part of theta to pi when (A+D)/2 > 1, giving cos(theta) = -|(A+D)/2|, which directly contradicts the defining relation cos(theta) = (A+D)/2; the correction adopts the standard principal-branch arccos so both real and complex regimes satisfy the definition consistently.}
\item \textbf{39.1} \,\textit{surface: Interface / return-contract mismatch} \,$\cdot$\, neutral-cleanup\\
{\footnotesize The original spec stated the return value as a flat tuple (A, B, C, D) of scalar elements, but downstream steps consume a 2x2 numpy array; the correction aligns the stated output contract with the actual required shape M = [[A, B], [C, D]].}
\end{itemize}

\subsection*{Problem 40 --- \texttt{Spliting\_Operator} \hfill \normalfont\textit{Mathematics, Computational Mechanics} (3 defects)}
{\footnotesize\emph{Original source:}~\citep{strang1968}}\\[1pt]
\begin{itemize}[leftmargin=1.1em,itemsep=1pt,topsep=1pt,parsep=0pt]
\item \textbf{40.3 (sub\_steps[2])} \,\textit{scientific: Invalid / non-reproducible target} \,$\cdot$\, too-strict/wrong\\
{\footnotesize The original spec used a nominal dx = dt/CFL = 0.05 in the diffusion discretization, but the actual grid spacing on x = linspace(-1, 1, N) is 2/(N-1) \textasciitilde{}= 0.04878, producing a \textasciitilde{}5\%-wrong diffusion operator; additionally, the time loop ran one step past T. The correction makes dx = x[1]-x[0] authoritative and integrates exactly round(T/dt) steps, regenerating the gold to match the self-consistent scheme.}
\item \textbf{40.2 (sub\_steps[1])} \,\textit{scientific: Unspecified convention} \,$\cdot$\, too-strict/wrong\\
{\footnotesize The original prompt did not state which operator receives the two dt/2 half-steps versus the single full-step dt in the Strang splitting, yet the gold is only reproduced (to \textasciitilde{}1e-16) by assigning the reaction term u\textasciicircum{}2 to the half-steps and the diffusion term alpha*u\_xx to the full step; the alternative reading fails all targets, so any correct implementation of the other convention is wrongly graded wrong.}
\item \textbf{40 (problem\_description\_main)} \,\textit{surface: Typo / formatting} \,$\cdot$\, neutral-cleanup\\
{\footnotesize The problem description contained spelling errors (`opeator', `spliting', `Eurler') and the phrasing `first order Strang splitting scheme' misattributes the first-order accuracy to the splitting itself rather than to the forward-Euler sub-steps; the correction fixes the typos and clarifies that the composite scheme is first order because each sub-step uses one forward-Euler update.}
\end{itemize}

\subsection*{Problem 41 --- \texttt{Structural\_stability\_in\_serial\_dilution} \hfill \normalfont\textit{Biology, Ecology} (2 defects)}
{\footnotesize\emph{Original source:}~\citep{wang2024}}\\[1pt]
\begin{itemize}[leftmargin=1.1em,itemsep=1pt,topsep=1pt,parsep=0pt]
\item \textbf{41.3} \,\textit{surface: Broken cross-reference} \,$\cdot$\, neutral-cleanup\\
{\footnotesize The step description instructs the solver to compute `the determinant of M' without ever defining M, leaving the intended matrix ambiguous; replacing M with `the normalized conversion matrix' identifies the correct operand.}
\item \textbf{41.3} \,\textit{scientific: Invalid / non-reproducible target} \,$\cdot$\, too-strict/wrong\\
{\footnotesize The third (R=4) test case carries an incorrect growth-rate matrix g, preference ranking pref, depletion order, and niche-length vector t, so the precomputed gold target is wrong; the corrected instance supplies consistent, recomputed values and re-generates the target against them.}
\end{itemize}

\subsection*{Problem 42 --- \texttt{The\_threshold\_current\_for\_multi\_quantum\_well\_lasers} \hfill \normalfont\textit{Material Science, Semiconductor Materials} (2 defects)}
{\footnotesize\emph{Original source:}~\citep{mcilroy1985}}\\[1pt]
\begin{itemize}[leftmargin=1.1em,itemsep=1pt,topsep=1pt,parsep=0pt]
\item \textbf{42.1, 42.2} \,\textit{scientific: Wrong gold (method / sign / symmetry)} \,$\cdot$\, too-strict/wrong\\
{\footnotesize The step-42.1 output is labeled `the gain coefficient G\_th at threshold condition' and step-42.2 re-labels the same quantity as `the modal gain', but the function computes and passes forward the single-well peak gain g\_w; both misnomers make the spec internally inconsistent and physically incorrect.}
\item \textbf{42.3} \,\textit{scientific: Wrong gold (method / sign / symmetry)} \,$\cdot$\, too-strict/wrong\\
{\footnotesize The background formula states eta*I\_th = J\_th*w*L, inserting a spurious injection-efficiency factor: eta already appears in the preceding relation eta*J\_th = n\_w*J\_w, so the total threshold current is simply I\_th = J\_th*w*L, and the extra eta causes any correctly derived current to be marked wrong.}
\end{itemize}

\subsection*{Problem 43 --- \texttt{two\_end\_fiber\_laser\_generator} \hfill \normalfont\textit{Physics, Optics} (3 defects)}
{\footnotesize\emph{Original source:}~\citep{kelson1998}}\\[1pt]
\begin{itemize}[leftmargin=1.1em,itemsep=1pt,topsep=1pt,parsep=0pt]
\item \textbf{43.1 and 43.3 (function\_header docstrings)} \,\textit{scientific: Wrong gold (method / sign / symmetry)} \,$\cdot$\, neutral-cleanup\\
{\footnotesize In steps 43.1 and 43.3, the docstrings label gamma\_s and gamma\_p as ``Gain coefficient,'' but these symbols denote the modal overlap (filling) factor---a dimensionless geometric quantity describing the spatial confinement of the optical mode, not a gain. The step-43.3 prompt body already uses ``overlap factors'' for the same variables, making the docstring internally contradictory. The relabeling to ``Overlap (filling) factor'' corrects the physical description without changing any computed quantity or grading target.}
\item \textbf{43.3 prompt + output specification (and problem\_io / function\_header return shape)} \,\textit{scientific: Invalid / non-reproducible target} \,$\cdot$\, too-strict/wrong\\
{\footnotesize The original prompt omitted the saturation-power formulas (P\_ssat, P\_psat), the BVP grid (100 equally spaced points on [0, L]), the initial guess, solver tolerance (1e-9), max\_nodes (100000), the output-power definition Pout = Ps(L)*(1-R2), and the output array shape (100,) sampled via sol.sol(z); any correct solver using different conventions would mismatch the precomputed gold and be graded wrong.}
\item \textbf{43.3 test\_cases / general\_tests (final test)} \,\textit{scientific: Non-discriminating test (physics)} \,$\cdot$\, too-lenient (test tightened)\\
{\footnotesize The original test asserted only that the inversion profile is monotone (nz[0] > nz[mid]), ignoring the returned output power Pout entirely; a solution producing a physically wrong Pout would pass, so the assertion is strengthened to also require Pout > 50 W.}
\end{itemize}

\subsection*{Problem 45 --- \texttt{finite\_difference\_heat\_equation} \hfill \normalfont\textit{Mathematics, Computational Mechanics} (6 defects)}
{\footnotesize\emph{Original source:}~\citep{courant1928}}\\[1pt]
\begin{itemize}[leftmargin=1.1em,itemsep=1pt,topsep=1pt,parsep=0pt]
\item \textbf{problem\_io; sub\_steps[0](step 45.1).function\_header; sub\_steps[3](step 45.4).function\_header} \,\textit{scientific: Wrong gold (method / sign / symmetry)} \,$\cdot$\, neutral-cleanup\\
{\footnotesize The spec names both material-temperature parameters `T1' (a duplicate) and mislabels `alpha2' as `heat conductivity' rather than `thermal diffusivity'; alpha2 is the diffusivity used throughout the update equations, so the wrong physical-quantity name contradicts the body text and misleads implementers.}
\item \textbf{sub\_steps[0](step 45.1).step\_description\_prompt} \,\textit{scientific: Unspecified convention} \,$\cdot$\, too-strict/wrong\\
{\footnotesize The original `second and third dimensions are x and y coordinates' is ambiguous about which axis is rows versus columns; the gold array shape (Nt, Ny, Nx) places y in the second dimension and x in the third, so the original wording can lead a correct solver to transpose the grid and fail all tests.}
\item \textbf{sub\_steps[1](step 45.2).step\_description\_prompt; sub\_steps[2](step 45.3).step\_description\_prompt} \,\textit{surface: Broken cross-reference} \,$\cdot$\, neutral-cleanup\\
{\footnotesize Both step 45.2 and 45.3 prompts contain a garbled cross-reference reading ``defined in .'' where the referent step number was dropped during authoring, leaving ``the temperature array defined in .'' with no resolvable antecedent. A model solving these steps cannot determine which prior step establishes the temperature array's structure, boundary layout, or indexing convention. The fix supplies the missing referent: ``defined in step 45.1.''}
\item \textbf{sub\_steps[2](step 45.3).step\_description\_prompt} \,\textit{scientific: Unspecified convention} \,$\cdot$\, too-strict/wrong\\
{\footnotesize The original `outward normal' sign convention for Neumann boundaries contradicts the gold implementation, which uses a positive-axis convention (T\_b = T\_in - N at the first row/column and T\_b = T\_in + N at the last); a solver following the outward-normal reading produces the wrong sign on three of four boundary faces.}
\item \textbf{sub\_steps[1](step 45.2).test\_cases; sub\_steps[2](step 45.3).test\_cases} \,\textit{surface: Trivially broken test} \,$\cdot$\, too-strict/wrong\\
{\footnotesize The test passes the boundary-condition array as a bare Python list rather than an np.array, so array-style indexing or dtype operations inside a correct implementation raise a TypeError; a fourth Neumann test case covering the last-column boundary was also absent, leaving the corrected positive-axis sign convention untested on the high-index side.}
\item \textbf{sub\_steps[3](step 45.4).step\_description\_prompt} \,\textit{scientific: Unspecified convention} \,$\cdot$\, too-strict/wrong\\
{\footnotesize The original spec names only `central difference' and the stability time-step, leaving the time-integration scheme (forward Euler), the diffusivity evaluation (pointwise, non-conservative form critical at the two-material interface), the grid spacing (dx = dy = 1), and the BC-application timing all unspecified; any of these choices can produce results that diverge from gold without the solver being physically wrong.}
\end{itemize}

\subsection*{Problem 46 --- \texttt{helium\_atom\_vmc} \hfill \normalfont\textit{Chemistry, Quantum Chemistry} (2 defects)}
{\footnotesize\emph{Original source:}~\citep{mcmillan1965}}\\[1pt]
\begin{itemize}[leftmargin=1.1em,itemsep=1pt,topsep=1pt,parsep=0pt]
\item \textbf{46.3} \,\textit{scientific: RNG-dependent grading} \,$\cdot$\, too-strict/wrong\\
{\footnotesize The original test pins the exact RNG consumption order of a 2000-step seeded Metropolis walk via np.allclose against a precomputed trajectory, which rejects any correct implementation that uses a different but valid step size, batch size, or sweep order; the fix asserts instead that the time-average mean electron-nuclear distance satisfies |<r> - 3/(2*alpha)| < 0.02, the exact analytic value for the Slater trial wavefunction.}
\item \textbf{46.4} \,\textit{scientific: RNG-dependent grading} \,$\cdot$\, too-strict/wrong\\
{\footnotesize The original test uses cmp\_tuple\_or\_list exact-reproduction plus seeded interval checks that implicitly mandate a specific protocol (run Metropolis once, evaluate only the final configs, use ddof=0 standard error, do not average the chain), rejecting any valid time-averaging VMC estimator; the fix replaces this with 3-sigma statistical bounds against the analytic kinetic energy alpha\textasciicircum{}2, electron-nuclear potential -4*alpha, and electron-electron potential 5*alpha/8.}
\end{itemize}

\subsection*{Problem 48 --- \texttt{MEELS\_conversion} \hfill \normalfont\textit{Physics, Condensed Matter Physics} (4 defects)}
{\footnotesize\emph{Original source:}~\citep{vig2017}}\\[1pt]
\begin{itemize}[leftmargin=1.1em,itemsep=1pt,topsep=1pt,parsep=0pt]
\item \textbf{problem\_io + step 48.4 (step\_description\_prompt)} \,\textit{scientific: Unspecified convention} \,$\cdot$\, too-strict/wrong\\
{\footnotesize The spec names the output as chi''(omega) (the imaginary part of the density response), but the gold test expects the positive-omega slice to be positive, which is only true for -chi''; a correct chi'' implementation would fail the sign check.}
\item \textbf{step 48.2 (function\_header)} \,\textit{scientific: Wrong gold (method / sign / symmetry)} \,$\cdot$\, too-strict/wrong\\
{\footnotesize The function-header docstring labels the output V\_eff as having units of ``inverse of square angstrom'' (Angstrom\textasciicircum{}-2), but the Coulomb matrix element V\_eff is physically dimensioned in square angstroms (Angstrom\textasciicircum{}2)---off by inversion. A correct implementation that returns values in Angstrom\textasciicircum{}2 cannot be verified against this spec, because the stated unit directly contradicts the actual physical quantity being computed.}
\item \textbf{step 48.4 (step\_description\_prompt)} \,\textit{scientific: Unspecified convention} \,$\cdot$\, too-strict/wrong\\
{\footnotesize The antisymmetrization step never specifies how to evaluate S(-omega) on a non-uniform, non-symmetric grid; the gold uses linear interpolation, so a reasonable alternative such as nearest-neighbor lookup produces a different numerical result and fails the allclose check.}
\item \textbf{step 48.4 (test\_cases)} \,\textit{surface: Trivially broken test} \,$\cdot$\, too-strict/wrong\\
{\footnotesize The test applies a boolean numpy mask directly to the raw return value of chi\_cal; when the function returns a Python list (as the docstring specifies), this raises a TypeError and a correct solution is rejected without wrapping the return in np.array first.}
\end{itemize}

\subsection*{Problem 50 --- \texttt{Replica\_symmetry\_breaking} \hfill \normalfont\textit{Physics, Condensed Matter Physics} (7 defects)}
{\footnotesize\emph{Original source:}~\citep{sherrington1975}}\\[1pt]
\begin{itemize}[leftmargin=1.1em,itemsep=1pt,topsep=1pt,parsep=0pt]
\item \textbf{problem\_background\_main / sub\_steps[3](step 50.4).step\_background} \,\textit{scientific: Unspecified convention} \,$\cdot$\, neutral-cleanup\\
{\footnotesize The original spec defined J\_ij as i.i.d. N(0,1)/sqrt(N) without stating that J must be symmetric (J\_ij = J\_ji, required for the SK Hamiltonian to be well-defined) or that the diagonal is zero (no self-interaction); the correction makes the symmetric, zero-diagonal construction explicit.}
\item \textbf{sub\_steps[0](step 50.1).step\_description\_prompt} \,\textit{scientific: RNG-dependent grading} \,$\cdot$\, too-strict/wrong\\
{\footnotesize The constraint permitting only np.random.randint and np.random.rand is an artificial RNG-protocol requirement that pins the exact random-number stream rather than constraining physics, so any correct Metropolis implementation using other valid random draws fails the test.}
\item \textbf{sub\_steps[0](step 50.1).test\_cases} \,\textit{scientific: RNG-dependent grading} \,$\cdot$\, too-strict/wrong\\
{\footnotesize Asserting np.allclose(spins, target) on the final spin configuration requires bitwise reproduction of the gold MC RNG trajectory, not thermalization to a correct equilibrium; the replacement tests the mean equilibrated energy per spin against a physically derived reference (-0.4207 +/- 3*0.0223), accepting any correct equilibration.}
\item \textbf{sub\_steps[3](step 50.4).step\_description\_prompt} \,\textit{scientific: RNG-dependent grading} \,$\cdot$\, too-strict/wrong\\
{\footnotesize The constraint permitting only np.random.randn and np.random.choice pins the RNG stream so that seeded exact (RSB, mean, std) reproduction tests match bitwise; any correct implementation using other valid sampling functions fails, and the ddof=0 convention for the reported standard deviation was also left unspecified.}
\item \textbf{sub\_steps[3](step 50.4).test\_cases} \,\textit{scientific: RNG-dependent grading} \,$\cdot$\, too-strict/wrong\\
{\footnotesize Asserting exact reproduction of the (RSB, mean, std) triple via np.allclose passes only under bitwise RNG replication of the gold protocol; the replacement tests physically meaningful phase behaviour, checking that at high T=1.5 RSB is False with narrow overlaps and at low T=0.5 RSB is True with broad overlaps.}
\item \textbf{sub\_steps[2](step 50.3).step\_description\_prompt / sub\_steps[2](step 50.3).step\_background} \,\textit{scientific: Unspecified convention} \,$\cdot$\, neutral-cleanup\\
{\footnotesize The prompt gave no testable return criterion for replica-symmetry breaking, leaving `analyze the overall overlap distribution' entirely open-ended; the correction specifies the explicit threshold (std(overlaps) > 1.4/sqrt(N)) so solvers know exactly what Boolean to return.}
\item \textbf{sub\_steps[2](step 50.3).test\_cases} \,\textit{scientific: RNG-dependent grading} \,$\cdot$\, neutral-cleanup\\
{\footnotesize The test inputs for analyze\_rsb were drawn with np.random.normal without a seed, so the overlap standard deviation could straddle the RSB threshold across runs, making the Boolean assertion non-deterministically flaky; adding fixed seeds makes the test inputs reproducible without constraining the solver.}
\end{itemize}

\subsection*{Problem 52 --- \texttt{Shooting\_algo\_H\_atom} \hfill \normalfont\textit{Physics, Computational Physics} (6 defects)}
{\footnotesize\emph{Original source:}~\citep{haule509}}\\[1pt]
\begin{itemize}[leftmargin=1.1em,itemsep=1pt,topsep=1pt,parsep=0pt]
\item \textbf{52.1} \,\textit{scientific: Non-discriminating test (physics)} \,$\cdot$\, too-lenient (test tightened)\\
{\footnotesize All three original tests set u=0, making the second-derivative term (l(l+1)/r\textasciicircum{}2 - 2Z/r - En)*u identically zero regardless of the potential; a no-op implementation returning [y[1], 0] passes every case. Setting u0=1.0 in the first test makes the potential term nonzero and actually constrains the radial-Schrodinger physics.}
\item \textbf{52.2} \,\textit{surface: Interface / return-contract mismatch} \,$\cdot$\, too-strict/wrong\\
{\footnotesize The docstring declared the output ur as a scalar float and the input grid as `linespace', but every test target is a shape-(100,) float64 array passed over a logspace grid; a solver following the documented type and grid name fails both type checks and np.allclose against the array target.}
\item \textbf{52.2, 52.3} \,\textit{scientific: Invalid / non-reproducible target} \,$\cdot$\, too-strict/wrong\\
{\footnotesize Hydrogen bound states require negative energy; the original positive (scattering) energies produce inward-integrated profiles dominated by integrator noise, so the precomputed targets are unreproducible across integrator settings. Replacing En with negative bound-state values and narrowing the radius ranges yields well-conditioned integrals with consistent, regenerable targets.}
\item \textbf{52.4} \,\textit{surface: Broken cross-reference} \,$\cdot$\, too-strict/wrong\\
{\footnotesize The prompt referenced Shoot(En, R, l) with three arguments, but the actual function defined in step 52.3 requires a fourth parameter y0; calling the documented signature raises a TypeError. The prompt also described l as a `maximum' angular momentum, but FindBoundStates takes a single l value with the outer loop handled externally.}
\item \textbf{52.4} \,\textit{scientific: Over-tight tolerance} \,$\cdot$\, too-strict/wrong\\
{\footnotesize The shallowest bound-state energies from inward integration are sensitive to integrator step size at the level of \textasciitilde{}1e-7, so the default np.allclose tolerance (rtol=1e-5, atol=0) can reject a numerically correct solver. Adding atol=1e-5 admits this integrator-noise scatter while still rejecting physically wrong energies.}
\item \textbf{52.4} \,\textit{surface: Trivially broken test} \,$\cdot$\, too-lenient (test tightened)\\
{\footnotesize Using .any() in the assertion means the check passes whenever the first bound state has l=0, regardless of its energy value; for example, Bnd[0]=(0, 42.0) satisfies the original test. Replacing .any() with .all() requires both the angular momentum and the energy to match the target, making the test non-vacuous.}
\end{itemize}

\subsection*{Problem 53 --- \texttt{Stochastic\_Lotka\_Volterra} \hfill \normalfont\textit{Biology, Ecology} (4 defects)}
{\footnotesize\emph{Original source:}~\citep{gillespie1977}}\\[1pt]
\begin{itemize}[leftmargin=1.1em,itemsep=1pt,topsep=1pt,parsep=0pt]
\item \textbf{problem\_io; sub\_steps[2](step 53.3).function\_header; sub\_steps[3](step 53.4).function\_header} \,\textit{scientific: Unspecified convention} \,$\cdot$\, neutral-cleanup\\
{\footnotesize The phrase `rounded up to one decimal point' literally means ceiling, but the gold uses round-to-nearest at one decimal place; the ambiguity causes a correct implementation using round-to-nearest to be graded wrong.}
\item \textbf{sub\_steps[0](step 53.1).step\_background; sub\_steps[0](step 53.1).step\_description\_prompt} \,\textit{scientific: Wrong gold (method / sign / symmetry)} \,$\cdot$\, too-strict/wrong\\
{\footnotesize The original waiting-time PDF f=(1/a)exp(-dt/a) inverts the rate, giving mean a instead of the correct 1/a; additionally, omitting the sampling scale left solvers to guess whether to pass a or 1/a to NumPy's exponential, with the wrong choice shifting drawn times by roughly six orders of magnitude.}
\item \textbf{sub\_steps[1](step 53.2).step\_description\_prompt} \,\textit{scientific: Unspecified convention} \,$\cdot$\, too-strict/wrong\\
{\footnotesize The prompt is silent on whether an event whose time reaches or crosses T is recorded; the test compares exact time arrays, so including versus excluding the boundary event changes the output, marking a correct simulation wrong if it applies a different but physically valid boundary convention.}
\item \textbf{sub\_steps[3](step 53.4).test\_cases} \,\textit{scientific: Over-tight tolerance} \,$\cdot$\, too-strict/wrong\\
{\footnotesize Exact float equality on a spectral period estimate rounded to one decimal place rejects correct solutions that land on an adjacent decimal bin, including cases where ceiling and round-to-nearest differ by up to 0.1; replacing with abs difference < 0.11 admits only genuine one-decimal ambiguity while still rejecting wrong periods.}
\end{itemize}

\subsection*{Problem 54 --- \texttt{SUPG} \hfill \normalfont\textit{Mathematics, Computational Mechanics} (8 defects)}
{\footnotesize\emph{Original source:}~\citep{brooks1982}}\\[1pt]
\begin{itemize}[leftmargin=1.1em,itemsep=1pt,topsep=1pt,parsep=0pt]
\item \textbf{problem\_description\_main (+ sub\_steps[2] step 54.3 step\_description\_prompt)} \,\textit{scientific: Spec$\leftrightarrow$gold contradiction} \,$\cdot$\, too-strict/wrong\\
{\footnotesize The spec's tau formula uses the element Peclet number P\textasciicircum{}e = |a|h/2kappa, but the gold assembly is consistent only with a fixed constant P = |a| = 200 (independent of h and kappa=1); models obeying the stated formula compute a wrong tau and fail. Additionally, the 54.3 prompt said to add both Nitsche and SUPG terms, which double-counts the SUPG volume contribution already assembled in 54.2; the correction restricts 54.3 to Nitsche boundary terms only.}
\item \textbf{sub\_steps[1] (step 54.2) step\_description\_prompt} \,\textit{surface: Truncation / garble / dropped constant} \,$\cdot$\, too-strict/wrong\\
{\footnotesize The 54.2 step prompt omits the load-bearing constants a=200, kappa=1, and the tau formula; because the harness feeds each step in isolation (without the main problem description), a model has no source for these values and cannot assemble the correct matrix, forcing failure on missing information rather than capability.}
\item \textbf{sub\_steps[1] (step 54.2) step\_background} \,\textit{scientific: Wrong gold (method / sign / symmetry)} \,$\cdot$\, too-strict/wrong\\
{\footnotesize The 3-point Gauss-Legendre nodes are listed as +-3/5 = +-0.6, but the correct values are +-sqrt(3/5) \textasciitilde{} +-0.775; using the wrong nodes mis-integrates the degree-3 source term (12x\textasciicircum{}2 times a basis derivative), producing an RHS vector b that cannot match the gold.}
\item \textbf{problem\_io + sub\_steps[1] (54.2) / sub\_steps[2] (54.3) / sub\_steps[3] (54.4) function\_header} \,\textit{surface: Interface / return-contract mismatch} \,$\cdot$\, too-strict/wrong\\
{\footnotesize The spec documents the solution array and RHS vector as 1-dimensional arrays of size N+1 and M, but the gold returns 2-dimensional column vectors of shape (N+1,1) and (M,1); a solver returning the documented 1d shape fails the shape check in the L2Error test.}
\item \textbf{sub\_steps[0] (step 54.1) step\_description\_prompt} \,\textit{scientific: Unspecified convention} \,$\cdot$\, too-strict/wrong\\
{\footnotesize The basis-function prompt gives the piecewise formulas but never defines the node coordinates (x\_j = (j-1)h, 1-based index j) or states that each function returns 0 outside its support interval; without these, the piecewise assembly and downstream index arithmetic are ambiguous and inconsistent with the gold's zero-outside behavior.}
\item \textbf{sub\_steps[3] (step 54.4) step\_description\_prompt} \,\textit{surface: Truncation / garble / dropped constant} \,$\cdot$\, neutral-cleanup\\
{\footnotesize The 54.4 prompt reads `using , and functions' with the dependency names dropped, leaving the instruction grammatically incomplete and uninterpretable; restoring the missing names (basis, assemble, stabilization) is a text-integrity fix with no change to difficulty.}
\item \textbf{sub\_steps[3] (step 54.4) test\_cases + general\_tests (L2Error helper)} \,\textit{surface: Trivially broken test} \,$\cdot$\, too-strict/wrong\\
{\footnotesize The L2Error test helper loops over range(1,N), skipping the last element, and reconstructs u\_h as sol[e]*basis(e,..,etype=1) + sol[e+1]*basis(e+1,..,etype=1), which uses wrong node offsets and applies etype=1 to both nodes; the correct reconstruction is sol[e-1]*basis(e,..,etype=2) + sol[e]*basis(e+1,..,etype=1) over range(1,N+1), so the original test measured error against a wrong approximate solution.}
\item \textbf{sub\_steps[1] (step 54.2) step\_background} \,\textit{scientific: Unspecified convention} \,$\cdot$\, too-strict/wrong\\
{\footnotesize The 54.2 background describes A as only the convection-diffusion bilinear form and b as only the int 12x\textasciicircum{}2 omega term, omitting the SUPG volume contributions (int tau a omega\_,x (a u\_h,x) dx for A and int tau a omega\_,x (12x\textasciicircum{}2) dx for b) that the gold assembly includes; a model following the background alone assembles a wrong-per-gold matrix and right-hand side.}
\end{itemize}

\subsection*{Problem 55 --- \texttt{Swift\_Hohenberg} \hfill \normalfont\textit{Physics, Condensed Matter Physics} (6 defects)}
{\footnotesize\emph{Original source:}~\citep{swift1977}}\\[1pt]
\begin{itemize}[leftmargin=1.1em,itemsep=1pt,topsep=1pt,parsep=0pt]
\item \textbf{55.1} \,\textit{scientific: Unspecified convention} \,$\cdot$\, too-strict/wrong\\
{\footnotesize The original spec describes the split-step scheme only qualitatively, omitting the explicit-Euler nonlinear update u\_1 = u\_N + dt(eps*u\_N - u\_N - u\_N\textasciicircum{}3), the exact k-space propagator exp[dt(2k\textasciicircum{}2/q0\textasciicircum{}2 - k\textasciicircum{}4/q0\textasciicircum{}4)], the dx=1 wavenumber convention, the real-part projection, and the integer step count, so any legitimate alternative discretization fails the precomputed targets even if physically correct.}
\item \textbf{55.2} \,\textit{scientific: Unspecified convention} \,$\cdot$\, too-strict/wrong\\
{\footnotesize The spec names Kx, Ky, and Sk as returned arrays without specifying the meshgrid orientation (indexing='ij' vs `xy' transposes the result) or that Sk = |fft2(u)|\textasciicircum{}2 with no normalization, fftshift-centered; a correct solver choosing the other convention fails element-wise comparison against the precomputed target.}
\item \textbf{55.3} \,\textit{scientific: Unspecified convention} \,$\cdot$\, too-strict/wrong\\
{\footnotesize The original peak-detection recipe leaves the bin count, range, peak-finding method, and proximity tolerance (`some reasonable tolerance threshold') unspecified, making the precomputed target unreproducible; the correction pins N//2 equal-width bins over [0, max k\_r], bin-center reporting, scipy.signal.find\_peaks with a height threshold, and a 0.5*q0 proximity window.}
\item \textbf{55.3} \,\textit{surface: Broken cross-reference} \,$\cdot$\, neutral-cleanup\\
{\footnotesize The spec lists a `narrowness' criterion for peak\_found\_near\_q0 that the actual detection algorithm (proximity + height only) never applies, and uses the inconsistent output name peak\_near\_q0\_location instead of peak\_location\_near\_q0; the correction removes the phantom criterion and aligns the identifier.}
\item \textbf{55.4} \,\textit{surface: Broken cross-reference} \,$\cdot$\, neutral-cleanup\\
{\footnotesize The `set to 0 if no stripe is formed' clause was attached to the min\_height input parameter, where it is meaningless (min\_height is a caller-supplied threshold, not a return value); the correction moves it to the stripe\_mode output, where it correctly describes the no-pattern return convention.}
\item \textbf{55.4} \,\textit{scientific: Over-tight tolerance} \,$\cdot$\, too-strict/wrong\\
{\footnotesize The test compared the full output tuple including the chaotic, RNG- and discretization-sensitive u field and raw Sk to precomputed bit-exact values; any correct solver using a legitimate scheme variant fails even when the physically meaningful observables (if\_form\_stripes and stripe\_mode) match, so the correction replaces full-tuple comparison with shape checks on u/Sk plus allclose on stripe\_mode and equality on if\_form\_stripes.}
\end{itemize}

\subsection*{Problem 57 --- \texttt{1D\_harmonic\_oscillator\_numerov\_shooting} \hfill \normalfont\textit{Physics, Computational Physics} (3 defects)}
{\footnotesize\emph{Original source:}~\citep{noumerov1924}}\\[1pt]
\begin{itemize}[leftmargin=1.1em,itemsep=1pt,topsep=1pt,parsep=0pt]
\item \textbf{57.2} \,\textit{scientific: Unspecified convention} \,$\cdot$\, too-strict/wrong\\
{\footnotesize The original spec omits the Numerov seed values (u0 = u\_b, u1 = u\_b + step*up\_b) and does not say whether step carries its sign, so any implementation using a different seed or abs(step) is rejected by the gold tests despite being a valid Numerov integrator.}
\item \textbf{57.4} \,\textit{scientific: Unspecified convention} \,$\cdot$\, too-strict/wrong\\
{\footnotesize The original spec says to count sign changes between consecutive elements but leaves zero-element handling undefined; the gold uses a product < 0 rule (zeros are sign-neutral), so an implementation treating a zero as a sign-flip returns the wrong node count and fails the test.}
\item \textbf{57.5} \,\textit{scientific: Unspecified convention} \,$\cdot$\, too-strict/wrong\\
{\footnotesize The spec does not state that bound-state energies are taken as the first trial energy where the node count changes, with no further refinement; the gold stores coarse first-crossing values on the energy grid (e.g. n=0 energy \textasciitilde{}1.0289 rather than the exact eigenvalue 1.0), so a more accurate bisection-refined result is graded wrong.}
\end{itemize}

\subsection*{Problem 58 --- \texttt{Tolman\_Oppenheimer\_Volkoff\_star} \hfill \normalfont\textit{Physics, Astrophysics} (6 defects)}
{\footnotesize\emph{Original source:}~\citep{oppenheimer1939}}\\[1pt]
\begin{itemize}[leftmargin=1.1em,itemsep=1pt,topsep=1pt,parsep=0pt]
\item \textbf{58.2} \,\textit{scientific: Wrong gold (method / sign / symmetry)} \,$\cdot$\, too-strict/wrong\\
{\footnotesize The docstring names the return value `eps: the specific internal energy' but the function eos\_rho\_from\_press computes and returns density, so a docstring-following solver returns the wrong physical quantity and fails every test.}
\item \textbf{58.3} \,\textit{surface: Broken cross-reference} \,$\cdot$\, too-strict/wrong\\
{\footnotesize Step 58.3 describes eos\_eps\_from\_press but tells the solver to take density rho as input (it takes pressure) and invokes the wrong function eos\_rho\_from\_press in every test, making a correct implementation of the intended function untestable.}
\item \textbf{58.4} \,\textit{surface: Broken cross-reference} \,$\cdot$\, neutral-cleanup\\
{\footnotesize The step 58.4 prompt refers to helpers eps\_from\_press and rho\_from\_press, which are undefined; the functions established in prior steps are eos\_eps\_from\_press and eos\_rho\_from\_press, so the broken names prevent a solver from locating the required helpers.}
\item \textbf{58.4} \,\textit{scientific: Wrong gold (method / sign / symmetry)} \,$\cdot$\, too-strict/wrong\\
{\footnotesize The background states the mass-continuity ODE as dm/dr = 4 pi r\textasciicircum{}3 mu, but the correct GR relation is dm/dr = 4 pi r\textasciicircum{}2 mu; the erroneous r\textasciicircum{}3 exponent contradicts the gold integrand and marks a physically correct implementation wrong.}
\item \textbf{58.5} \,\textit{scientific: Unspecified convention} \,$\cdot$\, too-strict/wrong\\
{\footnotesize The original prompt omits the conventions required to reproduce gold: the uniform radial grid from r=0 to r\_max, the reference gauge phi(0)=0, the surface rule (first radius where P<=0), and the lapse formula sqrt(1-2M/R)*exp(-phi(R)); without these a correct-method solver cannot match the target output.}
\item \textbf{58.5} \,\textit{scientific: Over-tight tolerance} \,$\cdot$\, too-strict/wrong\\
{\footnotesize Default-tolerance np.allclose effectively demands exact agreement with one integrator's trajectory, rejecting correct solutions from other ODE solvers; additionally, the first test's grid (npoints=200, rmax=100000) is too coarse and numerically unstable to yield a well-resolved target, so both the tolerance and grid parameters suppress correct implementations.}
\end{itemize}

\subsection*{Problem 59 --- \texttt{VQE} \hfill \normalfont\textit{Physics, Quantum Information/Computing} (3 defects)}
{\footnotesize\emph{Original source:}~\citep{omalley2016}}\\[1pt]
\begin{itemize}[leftmargin=1.1em,itemsep=1pt,topsep=1pt,parsep=0pt]
\item \textbf{59.1} \,\textit{surface: Interface / return-contract mismatch} \,$\cdot$\, too-strict/wrong\\
{\footnotesize The scaffold's return statement references `Rz`, a z-specific variable name never defined for the general-axis function `rotation\_matrices(axis, theta)`, which builds `R`; any solution following the documented output variable raises NameError or returns nothing.}
\item \textbf{59.2} \,\textit{scientific: Non-discriminating test (physics)} \,$\cdot$\, too-strict/wrong\\
{\footnotesize The test compares the complex inner product vdot(ansatz\_o, ansatz\_c) directly against the real product of norms; a physically valid ansatz differing by a global phase e\textasciicircum{}(i*phi) satisfies the Cauchy-Schwarz equality in magnitude but not in real/imaginary parts, so np.isclose fails and the correct state is rejected. The fix uses np.abs on the inner product for phase invariance and adds an explicit unit-norm check.}
\item \textbf{59.3} \,\textit{scientific: Non-discriminating test (physics)} \,$\cdot$\, neutral-cleanup\\
{\footnotesize The original third test case (U = CNOT21 * (H S\_dag x H S\_dag), psi = |1,-1j> x |1,-1j>/2) gives a Z1 expectation coinciding with earlier cases and adds no new coverage; it is replaced with U = I, psi = |10>, which exercises Z1 = -1 (first qubit in state |1>) as a distinct branch. A correct implementation passes both forms, so this is a test-coverage redesign with no effect on pass/fail direction.}
\end{itemize}

\subsection*{Problem 60 --- \texttt{Widom\_particle\_insertion} \hfill \normalfont\textit{Material Science, Molecular Modeling} (7 defects)}
{\footnotesize\emph{Original source:}~\citep{widom1963}}\\[1pt]
\begin{itemize}[leftmargin=1.1em,itemsep=1pt,topsep=1pt,parsep=0pt]
\item \textbf{60.2} \,\textit{scientific: Wrong gold (method / sign / symmetry)} \,$\cdot$\, too-strict/wrong\\
{\footnotesize The spec labels the potential `truncated and shifted' (V\textasciicircum{}\{tr-sh\}) but the case formula applies no shift: it returns V\_LJ(r) for r < r\_c and 0 otherwise. A true shifted potential subtracts V\_LJ(r\_c) from every pair, changing the insertion energy; the contradictory label misleads solvers into implementing the wrong variant.}
\item \textbf{60.3} \,\textit{scientific: RNG-dependent grading} \,$\cdot$\, too-strict/wrong\\
{\footnotesize The original tests assert exact float values for a single random insertion (e.g. 1.0185805629757558), which only pass if the solver reproduces the identical legacy-RNG call protocol. The correction checks the statistically correct ensemble observable mean(exp(-beta*dU)) over 20000 insertions with tolerance 0.02, accepting any correct RNG implementation.}
\item \textbf{60.3} \,\textit{surface: Typo / formatting} \,$\cdot$\, neutral-cleanup\\
{\footnotesize The original return description vaguely named `the test particle' without specifying that exactly one particle is inserted at a uniformly random position per call; the correction makes the single-insertion contract explicit. This is a wording clarification with no effect on difficulty.}
\item \textbf{60.4} \,\textit{scientific: Unspecified convention} \,$\cdot$\, too-strict/wrong\\
{\footnotesize The original spec described only `a regular grid ... properly positioned' without pinning cell-center offset, grid rounding (n = ceil(N\textasciicircum{}(1/3))), enumeration order (k fastest, then j, then i), or that the full n\textasciicircum{}3 sites are returned even when n\textasciicircum{}3 > N. Any of these ambiguities cause a correct but differently-gridded solver to fail the reference tests.}
\item \textbf{60.5} \,\textit{surface: Interface / return-contract mismatch} \,$\cdot$\, too-strict/wrong\\
{\footnotesize The prompt listed `positions' and `L' as parameters of the MC function, but the actual signature takes N and rho; neither `positions' nor `L' exists as an argument. A solver following the prose would construct the wrong call, failing before any physics is evaluated.}
\item \textbf{60.5} \,\textit{scientific: Unspecified convention} \,$\cdot$\, too-strict/wrong\\
{\footnotesize The return spec described `corrected energy' and `extended chemical potential' without giving the LJ long-range tail formulas (u\_tail and mu\_tail). Without the explicit (8/3)*pi*rho*epsilon*sigma\textasciicircum{}3*((1/3)*(sigma/r\_c)\textasciicircum{}9 - (sigma/r\_c)\textasciicircum{}3) expressions, a correct but differently-convention solver cannot match the reference mu\_ext values the test checks.}
\item \textbf{60.5} \,\textit{scientific: Invalid / non-reproducible target} \,$\cdot$\, too-strict/wrong\\
{\footnotesize The test included reference mu\_ext values for rho = 0.8 (16.14) and rho = 0.9 (54.56) produced by a short MC run that has not converged at those high densities; a correct implementation's stochastic estimate for these points falls outside the 0.1 mean-relative-error gate, causing spurious failures. Dropping the two unconverged density points fixes this.}
\end{itemize}

\subsection*{Problem 61 --- \texttt{Xray\_conversion\_I} \hfill \normalfont\textit{Physics, Condensed Matter Physics} (3 defects)}
{\footnotesize\emph{Original source:}~\citep{busing1967}}\\[1pt]
\begin{itemize}[leftmargin=1.1em,itemsep=1pt,topsep=1pt,parsep=0pt]
\item \textbf{61.3} \,\textit{scientific: Wrong gold (method / sign / symmetry)} \,$\cdot$\, neutral-cleanup\\
{\footnotesize The docstring for parameter z\_s in step 61.3 labels it as a step size in the phi rotation, but the function implements a theta-rotation step, and the rest of the problem (including step 61.5) consistently uses z\_s as a theta step. The mislabeled axis creates a direct contradiction between the documented parameter semantics and the actual rotation convention, which can cause implementers to parameterize or interpret the rotation incorrectly. Correcting phi to theta restores internal consistency with no change to task difficulty.}
\item \textbf{61.5} \,\textit{surface: Broken cross-reference} \,$\cdot$\, neutral-cleanup\\
{\footnotesize Step 61.5's background instructions contained two unresolved cross-references (`Employ step to calculate' and `G matrix obtained from step') with no step number supplied, and dropped the leading list item `1.', leaving the two-step procedure unnavigable; corrected by naming step 2 (q\_cal) and identifying G as the goniometer rotation matrix.}
\item \textbf{61.5} \,\textit{surface: Truncation / garble / dropped constant} \,$\cdot$\, neutral-cleanup\\
{\footnotesize In the step 61.5 function header, the docstring parameter description for z\_s originally read ``step size in the heta rotation'': the backslash-t of ``theta'' was consumed as a tab escape, leaving a run of whitespace followed by the fragment ``heta''. The rotation axis is thus unnamed and the parameter description is unreadable. The fix restores the intended token to a legible ``theta'', a neutral textual correction that does not alter the task or grading difficulty.}
\end{itemize}

\subsection*{Problem 62 --- \texttt{dmrg} \hfill \normalfont\textit{Physics, Condensed Matter Physics} (4 defects)}
{\footnotesize\emph{Original source:}~\citep{white1992}}\\[1pt]
\begin{itemize}[leftmargin=1.1em,itemsep=1pt,topsep=1pt,parsep=0pt]
\item \textbf{62.1} \,\textit{surface: Interface / return-contract mismatch} \,$\cdot$\, too-strict/wrong\\
{\footnotesize The scaffold defined only `class EnlargedBlock` with its `\_\_init\_\_` and `print\_all` methods duplicated, while `class Block` (required for single-site blocks) was entirely absent, making the provided header syntactically and logically unusable.}
\item \textbf{62.3} \,\textit{surface: Broken cross-reference} \,$\cdot$\, neutral-cleanup\\
{\footnotesize Three cross-references in the problem text were left blank (``from step ,'', ``defined in step'', ``as described in step .''), forcing solvers to guess which prior steps define the required operators and Hamiltonian; the correction fills in the correct step numbers (62.2, 62.4).}
\item \textbf{62.5} \,\textit{scientific: Non-discriminating test (physics)} \,$\cdot$\, too-strict/wrong\\
{\footnotesize The DMRG truncation step operates on a degenerate SU(2) multiplet, so the eigenbasis within the degenerate subspace is defined only up to an arbitrary orthogonal rotation; comparing raw operator-matrix entries rejects correct implementations that happen to use a different valid gauge. The fix checks basis-invariant quantities (eigenvalues of H, singular values of conn\_Sz and conn\_Sp) instead.}
\item \textbf{62.6} \,\textit{surface: Trivially broken test} \,$\cdot$\, neutral-cleanup\\
{\footnotesize The test suite for step 62.6 listed the run\_dmrg(block, 100, 100, model\_d) call twice---once as case 3 and again as a trailing case 6---producing a redundant duplicate entry with no distinct parameter variation. The corrected suite retains five unique test cases (varying m0 and m) and drops the repeated case, eliminating unnecessary redundancy without changing which solutions pass or fail.}
\end{itemize}

\subsection*{Problem 63 --- \texttt{Estimating\_Stock\_Option\_Price} \hfill \normalfont\textit{Mathematics, Computational Finance} (7 defects)}
{\footnotesize\emph{Original source:}~\citep{schwartz1977}}\\[1pt]
\begin{itemize}[leftmargin=1.1em,itemsep=1pt,topsep=1pt,parsep=0pt]
\item \textbf{63.2} \,\textit{scientific: Wrong gold (method / sign / symmetry)} \,$\cdot$\, too-strict/wrong\\
{\footnotesize The background equation gives the expiry payoff as V(t\_max, p) = max(K-s, 0), which is the European put payoff, but the problem prices a European call whose correct terminal condition is V(t\_max, p) = max(s-K, 0). This sign reversal in the boundary condition propagates through any backward-in-time PDE solver, producing option values that correspond to a put rather than a call. A correct implementation using the call payoff max(s-K, 0) cannot reproduce the wrong boundary specified in the background, so any such solution is graded as incorrect.}
\item \textbf{63.3} \,\textit{scientific: Wrong gold (method / sign / symmetry)} \,$\cdot$\, too-strict/wrong\\
{\footnotesize The explicit finite-difference update formula writes the c-coefficient term as c*V\textasciicircum{}n\_\{j-1\}, duplicating the left-neighbour index; the standard three-point central-difference stencil requires c*V\textasciicircum{}n\_\{j+1\}. With the right neighbour absent the recurrence V\textasciicircum{}\{n+1\}\_j = a*V\textasciicircum{}n\_\{j-1\} + b*V\textasciicircum{}n\_j + c*V\textasciicircum{}n\_\{j-1\} is self-contradictory with the stencil stated just above it. Any reference solution or gold target generated from this formula propagates the wrong value at every time step, so a correct implementation of the intended scheme produces different numerical output and fails the comparison.}
\item \textbf{63.4} \,\textit{scientific: Wrong gold (method / sign / symmetry)} \,$\cdot$\, too-strict/wrong\\
{\footnotesize The docstring labels the tridiagonal matrix D as shape (N\_t-2)x(N\_t-2), but D operates on interior price nodes, making its correct shape (N\_p-2)x(N\_p-2); the wrong dimension label contradicts step 63.3's own description and misdirects any solver implementation.}
\item \textbf{63.1} \,\textit{scientific: Unspecified convention} \,$\cdot$\, too-strict/wrong\\
{\footnotesize The output description says only `np.linspace between p\_min and p\_max' without specifying that the grid is in log-price or that the bounds must stay in the given (descending) order; the test passes bounds already swapped, so an ascending linspace produces a different array and fails a correct implementation.}
\item \textbf{63.6} \,\textit{scientific: Non-discriminating test (physics)} \,$\cdot$\, too-strict/wrong\\
{\footnotesize The test sets S0=100 with min\_price=200 (= strike/5), placing the initial stock price below the grid's lower bound; interpolating V at log(S0) over a log-price grid that does not span that point is undefined, making the test ill-posed. The fix moves S0=400 in-range and specifies the floor time-index and log-grid linear-interpolation convention.}
\item \textbf{63.3} \,\textit{surface: Interface / return-contract mismatch} \,$\cdot$\, neutral-cleanup\\
{\footnotesize The original docstrings for steps 63.3 and 63.4 do not state that D is a scipy.sparse matrix, leaving the cross-step type contract implicit; since the gold implementation returns and consumes a sparse matrix, the omission can cause silent type mismatches in downstream steps.}
\item \textbf{63.2} \,\textit{scientific: Unspecified convention} \,$\cdot$\, too-strict/wrong\\
{\footnotesize The original prompt (`correctly puts boundary condition ... 2D array w.r.t. time and price') does not specify which axis corresponds to price versus time, the log-price form s=e\textasciicircum{}p used in the payoff and boundary rows, or the corner-precedence order; a correct solution with a transposed layout or different corner assignment would be wrongly rejected by the deterministic test.}
\end{itemize}

\subsection*{Problem 64 --- \texttt{GCMC} \hfill \normalfont\textit{Material Science, Molecular Modeling} (5 defects)}
{\footnotesize\emph{Original source:}~\citep{adams1975}}\\[1pt]
\begin{itemize}[leftmargin=1.1em,itemsep=1pt,topsep=1pt,parsep=0pt]
\item \textbf{64.2} \,\textit{surface: Trivially broken test} \,$\cdot$\, too-strict/wrong\\
{\footnotesize The dist() function returns a single float, but the test indexed it with [0]; a correct scalar return cannot be subscripted this way, so the assertion would crash or silently mis-compare against a correct solution.}
\item \textbf{64.4} \,\textit{surface: Interface / return-contract mismatch} \,$\cdot$\, neutral-cleanup\\
{\footnotesize The spec listed a phantom argument `integer i' absent from the function signature, omitted the required box length L, and misspelled `posistions'; the corrected description matches the actual E\_i(r, positions, L, sigma, epsilon) interface.}
\item \textbf{64.4} \,\textit{scientific: Unspecified convention} \,$\cdot$\, too-strict/wrong\\
{\footnotesize When two particles coincide under the minimum-image convention, the Lennard-Jones energy diverges; different correct implementations may return +inf or nan, so the original unspecified convention could cause a correct solution to fail the gold comparison.}
\item \textbf{64.6} \,\textit{scientific: Spec$\leftrightarrow$gold contradiction} \,$\cdot$\, too-strict/wrong\\
{\footnotesize The spec gave real SI argon constants (mass in kg, sigma in meters) alongside an incoherent instruction to `make J and s dimensionless', while the test cases run in Lennard-Jones reduced units (epsilon = sigma = m = 1, k\_B = 1); the contradictory spec forced incorrect dimensional choices that the gold rejects.}
\item \textbf{64.6} \,\textit{scientific: Over-tight tolerance} \,$\cdot$\, too-strict/wrong\\
{\footnotesize A finite Monte Carlo run (1e5 steps) produces an equilibrium particle-count average with inherent stochastic variance; a 1\% relative tolerance is too tight to reliably accept a correct GCMC implementation, whereas 5\% accommodates statistical fluctuations while still rejecting incorrect results.}
\end{itemize}

\subsection*{Problem 65 --- \texttt{GHZ\_protocol\_fidelity} \hfill \normalfont\textit{Physics, Quantum Information/Computing} (6 defects)}
{\footnotesize\emph{Original source:}~\citep{bennett1996}}\\[1pt]
\begin{itemize}[leftmargin=1.1em,itemsep=1pt,topsep=1pt,parsep=0pt]
\item \textbf{65.1} \,\textit{surface: Interface / return-contract mismatch} \,$\cdot$\, too-strict/wrong\\
{\footnotesize The function signature `def tensor()' accepts no arguments, so any call passing matrices raises a TypeError; adding *args makes the signature match the docstring's own requirement of `any number of nd arrays'.}
\item \textbf{65.1} \,\textit{surface: Interface / return-contract mismatch} \,$\cdot$\, too-strict/wrong\\
{\footnotesize The spec declares the Kronecker product always returns a 2-d array of floats, but tensoring two 1-d vectors yields a 1-d result; a correct implementation returning the right shape fails the shape check, and the dtype restriction wrongly rejects complex outputs.}
\item \textbf{65.2, 65.3, 65.6} \,\textit{surface: Interface / return-contract mismatch} \,$\cdot$\, too-strict/wrong\\
{\footnotesize Quantum-channel Kraus operators are generically complex-valued; labeling them `array of floats' in steps 65.2, 65.3, and 65.6 mislabels the type and causes type-based checks to reject correct complex implementations.}
\item \textbf{65.2} \,\textit{scientific: Unspecified convention} \,$\cdot$\, too-strict/wrong\\
{\footnotesize The `sys' parameter is described only as `list of subsystems' with no indexing convention stated; a solver choosing 0-based indices targets the wrong subsystem, producing a physically different result that fails gold tests.}
\item \textbf{65.3, 65.6} \,\textit{surface: Broken cross-reference} \,$\cdot$\, neutral-cleanup\\
{\footnotesize Both 65.3 and 65.6 prompts contain truncated cross-references (`apply\_channel function in .' and `the protocol in given by') with the step number missing, leaving solvers without the pointer to the required prerequisite function.}
\item \textbf{65.4} \,\textit{scientific: Wrong gold (method / sign / symmetry)} \,$\cdot$\, too-strict/wrong\\
{\footnotesize The original formula rho'' = V rho' V\textasciicircum{}dagger is dimensionally impossible for a two-party state: V is 2\textasciicircum{}n x 2, so the two-party unitary must be VV and the adjoint acts on the left; the correct formula is rho'' = (VV)\textasciicircum{}dagger rho' (VV), and any correct implementation of this is graded wrong against the broken spec.}
\end{itemize}

\subsection*{Problem 66 --- \texttt{kolmogorov\_crespi\_potential} \hfill \normalfont\textit{Material Science, Molecular Modeling} (5 defects)}
{\footnotesize\emph{Original source:}~\citep{kolmogorov2005}}\\[1pt]
\begin{itemize}[leftmargin=1.1em,itemsep=1pt,topsep=1pt,parsep=0pt]
\item \textbf{problem\_io + sub\_steps[5] (step 66.6) step\_description\_prompt + sub\_steps[5] (step 66.6) function\_header} \,\textit{scientific: Unspecified convention} \,$\cdot$\, too-strict/wrong\\
{\footnotesize The spec states only `KC potential energy' with no normalization, yet the test targets are per-atom averages (total double sum divided by Ntop + Nbot); a solver returning the raw double sum is wrongly rejected.}
\item \textbf{sub\_steps[5] (step 66.6) step\_description\_prompt} \,\textit{scientific: Spec$\leftrightarrow$gold contradiction} \,$\cdot$\, too-strict/wrong\\
{\footnotesize The prompt lists KC parameter values (e.g. z0=3.416084, C0=20.021583) that contradict the function-header defaults (z0=3.370060885645178, C0=21.78333851687074) used to compute the test targets, so a solver following the prompt's values is incorrectly rejected; additionally, `C = E-2' is a garbled literal.}
\item \textbf{sub\_steps[1] (step 66.2) step\_description\_prompt + sub\_steps[1] (step 66.2) function\_header} \,\textit{surface: Truncation / garble / dropped constant} \,$\cdot$\, too-strict/wrong\\
{\footnotesize The prompt and header declare the normal-vector return shape as (natoms,) instead of (natoms, 3), causing solvers that build the correct 3-D array to be rejected, and the instruction for correcting sign is truncated mid-token (`multiplying by -' instead of `multiplying by -1').}
\item \textbf{sub\_steps[5] (step 66.6) test\_cases + general\_tests} \,\textit{surface: Trivially broken test} \,$\cdot$\, too-lenient (test tightened)\\
{\footnotesize The misplaced parenthesis `np.abs(energy - energy\_ref < 2)' first evaluates the comparison to a bool then takes its absolute value, making the tolerance check one-sided (only requiring energy < energy\_ref + 2) and allowing arbitrarily large negative errors to pass; the corrected form `np.abs(energy - energy\_ref) < 2' enforces the intended two-sided bound.}
\item \textbf{sub\_steps[0] (step 66.1) step\_description\_prompt + sub\_steps[0] (step 66.1) function\_header} \,\textit{scientific: Unspecified convention} \,$\cdot$\, too-strict/wrong\\
{\footnotesize The spec leaves the graphene geometry construction entirely unspecified (no lattice vectors, no basis-atom ordering, no loop convention), yet the tests use np.allclose on absolute atom coordinates, so any correct geometry with a different-but-valid atom ordering is wrongly rejected.}
\end{itemize}

\subsection*{Problem 67 --- \texttt{LEG\_Dyson\_equation\_bulk} \hfill \normalfont\textit{Physics, Condensed Matter Physics} (3 defects)}
{\footnotesize\emph{Original source:}~\citep{jain1985}}\\[1pt]
\begin{itemize}[leftmargin=1.1em,itemsep=1pt,topsep=1pt,parsep=0pt]
\item \textbf{67.3, 67.4, 67.5, 67.6 (step\_description\_prompt and step\_background)} \,\textit{surface: Broken cross-reference} \,$\cdot$\, neutral-cleanup\\
{\footnotesize Five prompt and background fields contain dangling cross-references of the form `as described in step' or `introduced in step' with the target step number missing, leaving the reader unable to trace the cited dependency.}
\item \textbf{67.5 (step\_description\_prompt)} \,\textit{scientific: Wrong gold (method / sign / symmetry)} \,$\cdot$\, too-strict/wrong\\
{\footnotesize The step description labels the physical constant as hbar/m\_e, but hbar/m\_e carries dimensions of velocity times length (J.s / kg = m\textasciicircum{}2/s), which is dimensionally inconsistent with the stated units meV nm\textasciicircum{}2 (energy times area). The quoted numerical value 76.2 meV nm\textasciicircum{}2 is consistent with hbar\textasciicircum{}2/m\_e, not hbar/m\_e, and the test cases confirm this by using 7.62 meV nm\textasciicircum{}2 / m\_eff (the effective-mass scaling of hbar\textasciicircum{}2/m\_e) for Fermi energy and velocity calculations. Any solution that reads the spec literally and defines the constant as hbar/m\_e operates with a quantity of the wrong physical dimension, producing incorrect Fermi energies that cannot match the expected outputs.}
\item \textbf{67.6 (test\_cases and general\_tests)} \,\textit{scientific: Over-tight tolerance} \,$\cdot$\, too-strict/wrong\\
{\footnotesize The test compares an N=101 finite-matrix RPA numerical result against a closed-form analytic expression using a relative tolerance of 1e-15, which is effectively bit-exact and physically unachievable: a finite-difference construction of this size accumulates O(N * eps\_machine) rounding error, placing legitimate results near 1e-13 relative error or larger. Simultaneously, the np.allclose checks rely on default tolerances (rtol=1e-5, atol=1e-8) without explicit specification, making acceptance criteria brittle across environments. A correct solution matching the analytic result to any physically meaningful precision is rejected. The fix loosens the numeric-vs-analytic relative tolerance to 1e-8 and adds explicit atol=rtol=1e-10 to the allclose calls.}
\end{itemize}

\subsection*{Problem 68 --- \texttt{helium\_atom\_dmc} \hfill \normalfont\textit{Chemistry, Quantum Chemistry} (5 defects)}
{\footnotesize\emph{Original source:}~\citep{reynolds1982}}\\[1pt]
\begin{itemize}[leftmargin=1.1em,itemsep=1pt,topsep=1pt,parsep=0pt]
\item \textbf{68.3} \,\textit{scientific: Over-tight tolerance} \,$\cdot$\, too-strict/wrong\\
{\footnotesize The Laplacian tests assert np.allclose(RMSE, precomputed\_target), pinning the finite-difference RMSE to one exact numerical value; any correct implementation with a slightly different rounding path fails. The fix replaces this with the physically meaningful contract that the numerical-vs-analytic RMSE is small (< 1e-2).}
\item \textbf{68.5} \,\textit{scientific: RNG-dependent grading} \,$\cdot$\, too-strict/wrong\\
{\footnotesize The Metropolis test asserted exact agreement with a seeded final walker configuration, requiring bitwise reproduction of a specific RNG trajectory; since Metropolis equilibration to |psi|\textasciicircum{}2 is path-irrelevant, the only physically correct contract is convergence of the mean radius to the known stationary value 3/(2 alpha). The fix replaces the exact-position test with that statistical equilibration check and refactors the function to a single sweep so callers control the loop.}
\item \textbf{68.6} \,\textit{scientific: Unspecified convention} \,$\cdot$\, too-strict/wrong\\
{\footnotesize The docstring left drift\_new's normalization ambiguous (drift\_old explicitly stated `multiplied by dtau' but drift\_new did not) and gave no specification for the return value, so a solver could legitimately return the clipped min(1, ratio) acceptance probability and fail the gold test. The fix pins the convention that drift\_new is also multiplied by dtau and requires the raw uncapped Metropolis-Hastings ratio.}
\item \textbf{68.7} \,\textit{scientific: RNG-dependent grading} \,$\cdot$\, too-strict/wrong\\
{\footnotesize The original branching test asserts exact walker indices from a seeded np.random.multinomial draw, which implicitly mandates one specific RNG protocol; any correct alternative such as systematic or residual resampling produces different indices and fails despite satisfying the same physical contract. The true requirement for Diffusion Monte Carlo branching is only that the expected multiplicity of walker i is proportional to w\_i / mean(w); the seed-locked index check is a bitwise-reproduction artifact, not a physical correctness criterion. The fix replaces the exact-index assertion with a statistical frequency test over M=20000 draws using deliberately spread weights, so any unbiased resampling scheme passes while degenerate schemes (uniform sampling, w\textasciicircum{}2 weighting) correctly fail.}
\item \textbf{68.8} \,\textit{scientific: RNG-dependent grading} \,$\cdot$\, too-strict/wrong\\
{\footnotesize The four original tests seed the RNG and assert np.allclose on the full energy sequence returned by run\_dmc, pinning a specific RNG trajectory that depends on branch-selection order and warm-up details---implementation choices that are physically irrelevant to diffusion Monte Carlo. Any correct DMC implementation that uses a different branching scheme, walker ordering, or warm-up strategy will produce a different stochastic trajectory and fail the exact-sequence check even though its mixed estimator converges correctly to the helium ground-state energy (-2.903724 Ha). The fix replaces the trajectory-pinning assertions with a statistical test: the warm-up-dropped mean local energy must fall within 0.05 Ha of -2.903724 Ha, validated robust across seeds (|dev| \textasciitilde{} 0.008---0.010, dev + 3*SEM <= 0.022), and the prompt and return specification are rewritten to state the mixed-estimator contract instead of prescribing RNG-protocol details.}
\end{itemize}

\subsection*{Problem 69 --- \texttt{LEG\_Dyson\_equation\_semi\_infinite} \hfill \normalfont\textit{Physics, Condensed Matter Physics} (2 defects)}
{\footnotesize\emph{Original source:}~\citep{jain1985}}\\[1pt]
\begin{itemize}[leftmargin=1.1em,itemsep=1pt,topsep=1pt,parsep=0pt]
\item \textbf{69.1, 69.2, 69.3 (step\_description\_prompt)} \,\textit{surface: Typo / formatting} \,$\cdot$\, neutral-cleanup\\
{\footnotesize A stray editorial cross-reference annotation (`[duplicate LEG\_Dyson equation-bulk step]') from the problem-authoring process was left in the solver-facing prompt of steps 69.1, 69.2, and 69.3, polluting the spec with internal bookkeeping text irrelevant to the solver.}
\item \textbf{69.7 (step\_description\_prompt)} \,\textit{scientific: Unspecified convention} \,$\cdot$\, too-strict/wrong\\
{\footnotesize The Raman intensity formula in step 69.7 uses symbols alpha, V, delta, and kd without defining them or mapping them to function inputs (delta\_E, kd), so a correct implementation has no basis for choosing the right quantities and can be marked wrong due to the spec gap.}
\end{itemize}

\subsection*{Problem 71 --- \texttt{GADC\_rev\_coherent\_info} \hfill \normalfont\textit{Physics, Quantum Information/Computing} (11 defects)}
{\footnotesize\emph{Original source:}~\citep{khatri2020}}\\[1pt]
\begin{itemize}[leftmargin=1.1em,itemsep=1pt,topsep=1pt,parsep=0pt]
\item \textbf{71.1} \,\textit{surface: Interface / return-contract mismatch} \,$\cdot$\, too-strict/wrong\\
{\footnotesize The function signature `def ket(dim)' omits the `args' parameter that the docstring describes as the basis-vector index, making the header self-contradictory and the function uncallable with the required argument; the output shape (D, 1) was also left unspecified.}
\item \textbf{71.9} \,\textit{surface: Broken cross-reference} \,$\cdot$\, too-strict/wrong\\
{\footnotesize The step prompt contains three coupled errors: a broken cross-reference to a nonexistent `neg\_coh\_info in .' with an empty placeholder; the label `coherent information' instead of `channel reverse coherent information'; and the maximizing state amplitude sqrt(p)|00> + sqrt(1-p)|11>, which is swapped relative to the state defined in step 71.8, making the optimization target inconsistent.}
\item \textbf{71.9} \,\textit{scientific: Wrong gold (method / sign / symmetry)} \,$\cdot$\, too-strict/wrong\\
{\footnotesize The function GADC\_rev\_coh\_inf computes the channel reverse coherent information I\_R = S(B) - S(AB), which swaps the roles of output and environment relative to the standard coherent information I\_C = S(B) - S(E). The original docstring and return-variable name labeled the output as ``coherent information'' / channel\_coh\_info, misidentifying the quantity as I\_C---a distinct quantity with a different formula and generally different numerical value. A correct implementation of I\_R that returns the right number would be graded against a label claiming it should equal I\_C, causing mismatch and incorrect rejection.}
\item \textbf{71.2} \,\textit{surface: Interface / return-contract mismatch} \,$\cdot$\, too-strict/wrong\\
{\footnotesize The signature `def tensor()' accepts no arguments despite the docstring describing a variadic sequence of matrices; the function cannot receive any input. Output dimensionality behavior (1-D vs 2-D depending on input) was also unspecified.}
\item \textbf{71.3} \,\textit{surface: Interface / return-contract mismatch} \,$\cdot$\, too-strict/wrong\\
{\footnotesize The prompt used the same symbol `i' for both the subsystem index (which subsystem the channel acts on) and the Kraus-operator index K\_i, an ambiguous collision that makes the formula uninterpretable without guessing which `i' is meant where.}
\item \textbf{71.3} \,\textit{scientific: Unspecified convention} \,$\cdot$\, too-strict/wrong\\
{\footnotesize The permutation parameter `perm' was described only as `desired order', leaving the index base and mapping direction unstated; perm values like 2 and 3 are out of range for 0-based indexing, so only the 1-based convention with output-subsystem-k = input-subsystem-perm[k] reproduces the targets.}
\item \textbf{71.4} \,\textit{scientific: Unspecified convention} \,$\cdot$\, too-strict/wrong\\
{\footnotesize The `sys' parameter for apply\_channel did not state whether subsystem indices are 0-based or 1-based; the convention is load-bearing because a test uses sys=[2] on a 2-subsystem state, which is valid only under 1-based indexing.}
\item \textbf{71.5} \,\textit{scientific: Unspecified convention} \,$\cdot$\, too-strict/wrong\\
{\footnotesize The `sys' parameter for partial\_trace gave no indication of index base; the 1-based convention governs which subsystems are traced out, so leaving the base unstated renders the interface under-specified and grades implementations using 0-based indexing as wrong.}
\item \textbf{71.6} \,\textit{scientific: Unspecified convention} \,$\cdot$\, too-strict/wrong\\
{\footnotesize The prompt gave no formula for von Neumann entropy and left the logarithm base unspecified; entropy in nats versus bits differs by a factor of ln 2, so the missing `S(rho) = -tr(rho log\_2 rho)' formula and explicit `bits (log base 2)' pin are necessary to grade numerical outputs correctly.}
\item \textbf{71.8} \,\textit{scientific: Unspecified convention} \,$\cdot$\, too-strict/wrong\\
{\footnotesize The reverse coherent information formula I\_R(A|B) = S(A) - S(AB) references systems A and B without defining which physical qubit each label denotes; the assignment (A = reference qubit, B = channel output) is required to compute the correct marginals S(A) and S(AB).}
\item \textbf{71.8} \,\textit{scientific: Wrong gold (method / sign / symmetry)} \,$\cdot$\, too-strict/wrong\\
{\footnotesize The function neg\_rev\_coh\_info computes the reverse coherent information I\_R, but the original docstring and output variable were labeled `coherent information' / `neg\_I\_c', the distinct quantity I\_C; this mislabeling contradicts the function name and causes models implementing I\_R correctly to appear to return the wrong quantity.}
\end{itemize}

\subsection*{Problem 72 --- \texttt{ising\_model} \hfill \normalfont\textit{Physics, Condensed Matter Physics} (10 defects)}
{\footnotesize\emph{Original source:}~\citep{metropolis1953}}\\[1pt]
\begin{itemize}[leftmargin=1.1em,itemsep=1pt,topsep=1pt,parsep=0pt]
\item \textbf{72.1} \,\textit{surface: Typo / formatting} \,$\cdot$\, neutral-cleanup\\
{\footnotesize The documented neighbor order (left, above, right, below) is reordered to (above, right, below, left); since all four neighbors are summed, the change is cosmetic and leaves no gold or test behavior altered.}
\item \textbf{72.8} \,\textit{scientific: RNG-dependent grading} \,$\cdot$\, too-strict/wrong\\
{\footnotesize The original tests require exact per-temperature mag2 array reproduction under a fixed seed, which fails for any correct implementation that differs in RNG draw order; the replacement validates physics behavior (high magnetization at low T, low magnetization at high T, monotone gap exceeding 0.5) using sufficient equilibration sweeps.}
\item \textbf{72.9} \,\textit{scientific: Unspecified convention} \,$\cdot$\, too-strict/wrong\\
{\footnotesize The original spec says to return `the temperature at which the derivative is minimized,' but the discrete derivative np.diff(mag2)/np.diff(T) is defined on intervals rather than single temperatures, so the instruction is ambiguous about whether to return the left or right endpoint; the correction pins the convention as T\_list[argmin(...)] (the left endpoint of the steepest-drop interval).}
\item \textbf{72.9} \,\textit{scientific: RNG-dependent grading} \,$\cdot$\, too-strict/wrong\\
{\footnotesize The original tests call scan\_T with np.random.seed(0) and a small lattice (N=5, nsweeps=100), then assert np.allclose(calc\_transition(Ts, mag2), target): the precomputed target encodes a specific MC RNG trajectory, not the mathematical correctness of the derivative-argmin estimator. Any correct implementation that uses a different RNG state, sweep count, or Metropolis ordering produces a different mag2 array and fails the allclose comparison even though it correctly identifies the critical temperature. The fix replaces these RNG-coupled tests with a deterministic unit test on synthetic input ([1.0, 1.0, 0.2, 0.1] at T=[1, 2, 3, 4], where the argmin of the discrete derivative is unambiguously T=2.0) and a physics anchor asserting the recovered Tc lies within 0.2 of the exact 2D Ising value 2/ln(1+sqrt(2)) = 2.26919.}
\item \textbf{72.2} \,\textit{surface: Truncation / garble / dropped constant} \,$\cdot$\, too-strict/wrong\\
{\footnotesize The prompt is corrupted in two places: the spin value is truncated to `1 or -' (dropping the defining `-1') and the cross-reference to step 1 is garbled to `given in .', leaving both the spin alphabet and the available helper function unspecified.}
\item \textbf{72.3} \,\textit{surface: Truncation / garble / dropped constant} \,$\cdot$\, too-strict/wrong\\
{\footnotesize The prompts for energy-over-all-sites and total-magnetization both end mid-token with `either 1 or -', omitting the `1' that completes the spin alphabet \{+1,-1\} and leaving the lattice domain undefined.}
\item \textbf{72.5} \,\textit{scientific: Unspecified convention} \,$\cdot$\, too-strict/wrong\\
{\footnotesize The docstring labels the return `acceptance ratio' without specifying whether the Boltzmann factor is capped at 1; the correct Metropolis probability is min(1, exp(-beta*dH)), and the precomputed test target (54.5982) reflects an uncapped value rather than the correct capped value of 1.0.}
\item \textbf{72.6} \,\textit{surface: Typo / formatting} \,$\cdot$\, downstream change (not counted as a defect)\\
{\footnotesize The original prompt wording is minimal and clean; the change only naturalizes phrasing and restates the return value to align with the sweep-chain API redesign, not a defect in the original specification.}
\item \textbf{72.6} \,\textit{scientific: RNG-dependent grading} \,$\cdot$\, too-strict/wrong\\
{\footnotesize The original tests pin the exact post-sweep lattice state for a fixed seed, so they pass only if the model reproduces \name{}'s specific per-site RNG draw order and short-circuit policy; any correct Metropolis sweep that uses a different but valid traversal order fails, making the tests RNG-protocol over-specifications rather than physics checks.}
\item \textbf{72.7} \,\textit{surface: Interface / return-contract mismatch} \,$\cdot$\, too-strict/wrong\\
{\footnotesize The original docstring contradicts itself: the body says the function collects `iteration, temperature, energy, and magnetization\textasciicircum{}2 in a dataframe' while the Return field declares `mag2: (numpy array)', and neither version mentions thermalization; the spec<->return-type inconsistency makes the expected output ambiguous.}
\item \textbf{72.7} \,\textit{scientific: RNG-dependent grading} \,$\cdot$\, too-strict/wrong\\
{\footnotesize The original tests require np.allclose on the full per-sweep mag2 trajectory for a fixed seed, passing only when the model exactly reproduces \name{}'s RNG draw sequence across all sweeps; the replacement checks instead that the equilibrium mean mag2 lies within 3 blocking-error standard deviations of the expected value, accepting any physically correct Metropolis implementation.}
\item \textbf{72.8} \,\textit{surface: Interface / return-contract mismatch} \,$\cdot$\, downstream change (not counted as a defect)\\
{\footnotesize The scan\_T signature is updated from (Ts, N, nsweeps) to (Ts, N, n\_equil, n\_measure) to propagate the burn-in/measure split introduced in 72.7; the original scan\_T spec is sound and this is a mechanical API propagation, not an original defect.}
\end{itemize}

\subsection*{Problem 73 --- \texttt{Xray\_conversion\_II} \hfill \normalfont\textit{Physics, Condensed Matter Physics} (4 defects)}
{\footnotesize\emph{Original source:}~\citep{busing1967}}\\[1pt]
\begin{itemize}[leftmargin=1.1em,itemsep=1pt,topsep=1pt,parsep=0pt]
\item \textbf{73.3 / 73.5 / 73.8 / 73.9 (function\_header z\_s description)} \,\textit{scientific: Wrong gold (method / sign / symmetry)} \,$\cdot$\, neutral-cleanup\\
{\footnotesize The z\_s parameter (frame-stepping scan increment) was mislabeled as the phi rotation in four function headers; phi is a fixed diffractometer angle passed separately, so the scan axis is theta. The wrong label misleads implementations into confusing the two rotations.}
\item \textbf{73.2 (function\_header det\_d description)} \,\textit{scientific: Unspecified convention} \,$\cdot$\, neutral-cleanup\\
{\footnotesize The description `sample distance to the detector' is ambiguous between the perpendicular gap and the along-beam projection; the Q-calculation geometry requires the sample-to-beam-center distance measured along the incident beam (+x direction), which the corrected description specifies.}
\item \textbf{73.6 (step\_description\_prompt)} \,\textit{surface: Truncation / garble / dropped constant} \,$\cdot$\, neutral-cleanup\\
{\footnotesize The prompt defined only the scalar formula d* = 1/d, omitting the enumeration and grouping task: collect all (h,k,l) with d* at or below d*\_max and return a dictionary mapping each distinct d* value to its list of reflections, which is what the function signature and tests actually require.}
\item \textbf{73.6 / 73.7 / 73.8 / 73.9 + general\_tests (test\_cases)} \,\textit{scientific: Non-discriminating test (physics)} \,$\cdot$\, too-strict/wrong\\
{\footnotesize For a symmetric cell, powder-ring reflections, indexed hkl, and the U orientation matrix are defined only up to point-group symmetry (sign flips, permutations, signed-permutation gauge on U). The original exact comparators reject all symmetry-equivalent correct answers; the corrected tests use set-equality on reflections, sign/permutation-invariant |hkl| comparison, and require target.T @ result to be a signed permutation matrix.}
\end{itemize}

\subsection*{Problem 74 --- \texttt{Householder\_QR} \hfill \normalfont\textit{Mathematics, Numerical Linear Algebra} (1 defect)}
{\footnotesize\emph{Original source:}~\citep{householder1958}}\\[1pt]
\begin{itemize}[leftmargin=1.1em,itemsep=1pt,topsep=1pt,parsep=0pt]
\item \textbf{problem\_background\_main + sub\_steps[0] (step 74.1) step\_background} \,\textit{scientific: Unspecified convention} \,$\cdot$\, too-strict/wrong\\
{\footnotesize The spec stops after the v-direction formula and never states that all n reflectors k=1..n must be applied; in the square case m=n the final reflector acts on a 1x1 subcolumn as F=[-1], flipping the sign of R's last diagonal entry. A solver that stops at n-1 reflectors (a natural reading of the terse spec) produces R[2,2]=+1.498 instead of -1.498, and is wrongly rejected.}
\end{itemize}

\subsection*{Problem 75 --- \texttt{graphene\_tight\_binding} \hfill \normalfont\textit{Material Science, Semiconductor Materials} (4 defects)}
{\footnotesize\emph{Original source:}~\citep{moon2012}}\\[1pt]
\begin{itemize}[leftmargin=1.1em,itemsep=1pt,topsep=1pt,parsep=0pt]
\item \textbf{75.1} \,\textit{surface: Truncation / garble / dropped constant} \,$\cdot$\, too-strict/wrong\\
{\footnotesize The spec gives the decay constant b as a bare placeholder `(b,a.u.)\textasciicircum{}\{-1\}' with no numeric value; without b = 1.17 Bohr\textasciicircum{}\{-1\} the exponential hopping V\_pp \textasciitilde{} exp(-b * distance) cannot be evaluated, making correct solutions unreproducible. The unit string `b, a.u.' was also garbled, conflating the variable name with the unit.}
\item \textbf{75.2} \,\textit{surface: Broken cross-reference} \,$\cdot$\, neutral-cleanup\\
{\footnotesize The phrase `using the hopping evaluation from .' contains an empty reference target, leaving the required function unidentified; the correction names step 1 (hopping\_mk) as the source, resolving the dangling cross-reference.}
\item \textbf{75.2} \,\textit{scientific: Unspecified convention} \,$\cdot$\, too-strict/wrong\\
{\footnotesize The parameter description for di, dj does not state they are integer unit-cell indices (n1, n2) or how the displacement vector is formed; the convention d = basis[ai] - (basis[aj] + n1*latvecs[0] + n2*latvecs[1]) is load-bearing, and applying the shift to the wrong atom yields incorrect bond lengths that fail the gold test.}
\item \textbf{75.3} \,\textit{scientific: Unspecified convention} \,$\cdot$\, too-strict/wrong\\
{\footnotesize The prompt omits both the hop-inclusion cutoff (in-plane projected distance <= a0 = a/sqrt(3)) and the Bloch-sum formula H\_ij(k) = sum\_R (-t(d)) e\textasciicircum{}\{ik.d\}; without the nearest-neighbor cutoff the Hamiltonian includes spurious long-range hops, shifting eigenvalues from the gold values [-3.258, -2.117, 2.472, 2.903] and rejecting any correctly implemented general solver.}
\end{itemize}

\subsection*{Problem 76 --- \texttt{protein\_dna\_binding} \hfill \normalfont\textit{Biology, Genetics} (5 defects)}
{\footnotesize\emph{Original source:}~\citep{schneider1986}}\\[1pt]
\begin{itemize}[leftmargin=1.1em,itemsep=1pt,topsep=1pt,parsep=0pt]
\item \textbf{76.1 (problem\_description\_main; sub\_steps[0].step\_description\_prompt; sub\_steps[0].function\_header)} \,\textit{scientific: Unspecified convention} \,$\cdot$\, too-strict/wrong\\
{\footnotesize The original spec declares each PWM row a probability distribution (implying L1 normalization) yet types the output as an integer array, which is self-contradictory; the correction pins L2 (Euclidean) per-row normalization with a float dtype, giving a single coherent convention.}
\item \textbf{76.2 (sub\_steps[1].step\_background)} \,\textit{scientific: Unspecified convention} \,$\cdot$\, too-strict/wrong\\
{\footnotesize The KL-divergence formula uses the ambiguous symbol log, but the gold value (7.4987...) is reproduced only with the natural logarithm; using log base 2 or log base 10 yields a different result, so the unspecified base rejects correct solvers choosing a non-natural base.}
\item \textbf{76.4 (sub\_steps[3].function\_header)} \,\textit{scientific: Spec$\leftrightarrow$gold contradiction} \,$\cdot$\, too-strict/wrong\\
{\footnotesize The docstring promises an integer return value, but the tests assert None when no binding site is found; a correct implementation following the docstring returns an integer sentinel and fails those tests solely due to the spec-test contradiction.}
\item \textbf{76.4 (sub\_steps[3].test\_cases; general\_tests)} \,\textit{scientific: RNG-dependent grading} \,$\cdot$\, too-strict/wrong\\
{\footnotesize The scanner aggregates num\_runs=100 random samplings and returns the most-frequent position, so demanding exact equality to a single hard-coded target ties correctness to a specific RNG draw; the corrected test checks that the returned position falls within one motif-length of the known inserted position, rejecting wrong answers without over-specifying the stochastic procedure.}
\item \textbf{76.4 (sub\_steps[3].test\_cases; general\_tests)} \,\textit{surface: Trivially broken test} \,$\cdot$\, too-strict/wrong\\
{\footnotesize Tests for data2 and data3 generate DNA sequences with binding sites from those motifs but scan using load\_motif\_from\_df(data), a copy-paste mismatch; a correct scanner cannot detect a data2-inserted site by scanning with the data motif, so correct implementations fail these tests due to the wrong motif being loaded.}
\end{itemize}

\subsection*{Problem 77 --- \texttt{Berendsen\_thermostat} \hfill \normalfont\textit{Material Science, Molecular Modeling} (10 defects)}
{\footnotesize\emph{Original source:}~\citep{berendsen1984}}\\[1pt]
\begin{itemize}[leftmargin=1.1em,itemsep=1pt,topsep=1pt,parsep=0pt]
\item \textbf{77.1 (problem\_io) and 77.12 (velocityVerlet function\_header)} \,\textit{surface: Truncation / garble / dropped constant} \,$\cdot$\, neutral-cleanup\\
{\footnotesize The P\_target unit line is garbled by a copy-paste artifact---`units: bar.ostat. Set to 0 to deactivate, units: picoseconds.'---fusing parts of two different parameter descriptions; the correction reads simply `units: bar.'.}
\item \textbf{77.12 (velocityVerlet step\_description\_prompt)} \,\textit{scientific: Unspecified convention} \,$\cdot$\, too-strict/wrong\\
{\footnotesize The Berendsen barostat box-scaling factor requires an isothermal compressibility value, but the original spec omits it entirely; without the specific constant (gamma = 4.6e-5 bar\textasciicircum{}-1, water's value), any general correct implementation produces an unmatchable gold.}
\item \textbf{77.3 (dist\_v function\_header)} \,\textit{scientific: Unspecified convention} \,$\cdot$\, too-strict/wrong\\
{\footnotesize The dist\_v docstring declares a scalar float return (`minimum image distance') but the function actually returns the minimum-image displacement vector from r1 to r2; the wrong return type and missing direction convention cause downstream sign errors in force calculations.}
\item \textbf{77.5 (f\_ij function\_header)} \,\textit{surface: Interface / return-contract mismatch} \,$\cdot$\, too-strict/wrong\\
{\footnotesize The f\_ij docstring types argument r as a scalar float distance, but the function receives a 3D displacement vector; the `from particle i to particle j' direction is load-bearing for the sign of the computed force.}
\item \textbf{77.6 (E\_tail step\_background)} \,\textit{scientific: Wrong gold (method / sign / symmetry)} \,$\cdot$\, too-strict/wrong\\
{\footnotesize The Lennard-Jones energy-tail formula in the step background omits the required 1/V (V = L\textasciicircum{}3) normalization factor: the correct total tail correction is (8/3) pi N\textasciicircum{}2 eps sigma\textasciicircum{}3 [...] / V, but the original writes it without the /V and mislabels it ``per particle.'' Because the N\textasciicircum{}2 prefactor already counts all pair interactions, dividing by V converts the extensive sum to an intensive bulk correction; omitting it inflates the result by a factor of V (a factor of 1000 for a typical 10\textasciicircum{}3 simulation box). Any correct implementation that includes the 1/V factor therefore produces a value differing from the precomputed gold by three orders of magnitude and is incorrectly rejected.}
\item \textbf{77.7 (P\_tail step\_background)} \,\textit{scientific: Wrong gold (method / sign / symmetry)} \,$\cdot$\, too-strict/wrong\\
{\footnotesize The long-range pressure tail correction for a Lennard-Jones fluid is P\_tail = (16/3) * pi * (N\textasciicircum{}2/V\textasciicircum{}2) * epsilon * sigma\textasciicircum{}3 * [(2/3)(sigma/r\_c)\textasciicircum{}9 - (sigma/r\_c)\textasciicircum{}3], where the N\textasciicircum{}2/V\textasciicircum{}2 factor reflects the pair density squared integrated over the tail volume. The original formula omitted the 1/V\textasciicircum{}2 (V = L\textasciicircum{}3) factor entirely, making the precomputed gold too large by a factor of V\textasciicircum{}2. Any correct implementation that includes 1/V\textasciicircum{}2 therefore produces a pressure differing from the gold by orders of magnitude for typical simulation box sizes, causing the step to fail regardless of the solution's physical correctness.}
\item \textbf{77.7 (P\_tail function\_header)} \,\textit{scientific: Unspecified convention} \,$\cdot$\, too-strict/wrong\\
{\footnotesize P\_tail must return pressure in bar, but the unit conversion from zeptojoules per nm\textasciicircum{}3 to bar depends on knowing that L, sigma, and rc are in nanometers and epsilon is in zeptojoules; omitting these units from the docstring leaves implementations unable to derive the x10 conversion factor.}
\item \textbf{77.9 (temperature test\_cases)} \,\textit{surface: Trivially broken test} \,$\cdot$\, neutral-cleanup\\
{\footnotesize The N=1 temperature test passes a 1D shape-(3,) velocity array where the function's contract is shape-(N,3); the fix reshapes it to (1,3) for contract consistency, though the numerical result is unchanged since np.sum(v**2) is shape-invariant here.}
\item \textbf{77.10 (pressure step\_background)} \,\textit{scientific: Unspecified convention} \,$\cdot$\, too-strict/wrong\\
{\footnotesize The virial-pressure formula defines r\_ij as `displacement from particle i to j' (= r\_j - r\_i), but the correct virial convention requires r\_ij = r\_i - r\_j; using the wrong sign flips the sign of the virial pressure, turning a correct +8.851 result into -8.851.}
\item \textbf{77.10 (pressure test\_cases) and 77.11 (forces test\_cases)} \,\textit{surface: Trivially broken test} \,$\cdot$\, too-strict/wrong\\
{\footnotesize In the second test case of both the pressure (77.10) and forces (77.11) steps, N is declared as 2 but the positions array has three rows, making the test ill-defined; the fix removes the spurious third row so the array shape matches N.}
\end{itemize}

\subsection*{Problem 79 --- \texttt{Nose\_Hoover\_chain\_thermostat} \hfill \normalfont\textit{Material Science, Molecular Modeling} (8 defects)}
{\footnotesize\emph{Original source:}~\citep{martyna1996}}\\[1pt]
\begin{itemize}[leftmargin=1.1em,itemsep=1pt,topsep=1pt,parsep=0pt]
\item \textbf{problem\_io / sub\_steps[3] (step 79.4) function\_header} \,\textit{surface: Interface / return-contract mismatch} \,$\cdot$\, too-strict/wrong\\
{\footnotesize The spec declared position and velocity trajectories as shape (nsteps, 1), but the gold comparison expects 1-D arrays of shape (nsteps,); a solver following the stated shape literally produces a 2-D output and is wrongly rejected.}
\item \textbf{sub\_steps[3] (step 79.4) function\_header} \,\textit{scientific: Unspecified convention} \,$\cdot$\, too-strict/wrong\\
{\footnotesize The original header gave no convention for trajectory indexing, leaving ambiguous whether the first entry is the initial state or the state after one step; the correction pins the record-then-step convention (x[0]=x0, v[0]=v0), which is load-bearing for matching the gold trajectory.}
\item \textbf{sub\_steps[1] (step 79.2) and sub\_steps[2] (step 79.3) function\_header} \,\textit{surface: Interface / return-contract mismatch} \,$\cdot$\, too-strict/wrong\\
{\footnotesize The Nose-Hoover chain of length M has per-link forces G\_i, thermostat velocities v\_xi\_i, and positions xi\_i, each a vector of shape (M,); declaring them as scalar float contradicts the multi-link algorithm and causes a correct array-based implementation to be rejected.}
\item \textbf{sub\_steps[1] (step 79.2) and sub\_steps[2] (step 79.3) function\_header} \,\textit{scientific: Unspecified convention} \,$\cdot$\, too-strict/wrong\\
{\footnotesize The original docstring described dt only as `the integration time step', concealing that nhc\_step advances the chain by half its argument; solvers following the Liouville framing (exp(iL w\_j dt/2)) passed w\_j*dt/2 instead of w\_j*dt, producing a factor-of-2 error in integration time.}
\item \textbf{sub\_steps[1] (step 79.2) step\_description\_prompt} \,\textit{scientific: Unspecified convention} \,$\cdot$\, too-strict/wrong\\
{\footnotesize The prompt used k\_B and Q\_k in the G\_1/G\_k formulas without fixing their numerical values; multiple choices (e.g. Q=1 or k\_B=2) each fail the targets, while only k\_B=1 and Q\_k=k\_B*T/omega\textasciicircum{}2 reproduce the gold, making the step ill-posed without these conventions.}
\item \textbf{sub\_steps[2] (step 79.3) step\_description\_prompt} \,\textit{scientific: Unspecified convention} \,$\cdot$\, too-strict/wrong\\
{\footnotesize Naming `Yoshida's fourth-order method' is insufficient because multiple 4th-order composition schemes exist with different coefficients; without pinning the triple-jump weights w1=1/(2-2\textasciicircum{}(1/3)), w2=1-2*w1, the gold trajectory is ambiguous and correct implementations using a different valid scheme are wrongly rejected.}
\item \textbf{sub\_steps[0] (step 79.1) test\_cases} \,\textit{scientific: Non-discriminating test (physics)} \,$\cdot$\, too-lenient (test tightened)\\
{\footnotesize All three original Verlet test cases set x0=0.0, so an implementation that mishandles the initial-position term in the position update passes regardless; adding a case with x0=1.0 exercises the position dependence and closes this false-pass hole.}
\item \textbf{sub\_steps[3] (step 79.4) test\_cases / general\_tests} \,\textit{scientific: Over-tight tolerance} \,$\cdot$\, too-strict/wrong\\
{\footnotesize The Nose-Hoover chain (NHC) integrator is chaotic: for M=2 chains, trajectories computed with N=20000 and N=40000 steps accumulate exponentially diverging floating-point rounding errors across any two correct-but-not-bit-identical implementations. The original test compared full trajectories element-wise against a precomputed gold, so a correct solver whose intermediate floats differ by machine-epsilon amounts is rejected at step counts far beyond the Lyapunov time. Reducing the M=2 cases to N=500 and N=1000 keeps the integration within the regime where correct implementations remain mutually reproducible under exact comparison, eliminating false failures without weakening coverage of the two-thermostat chain.}
\end{itemize}

\subsection*{Problem 80 --- \texttt{Anderson\_thermostat} \hfill \normalfont\textit{Material Science, Molecular Modeling} (5 defects)}
{\footnotesize\emph{Original source:}~\citep{andersen1980}}\\[1pt]
\begin{itemize}[leftmargin=1.1em,itemsep=1pt,topsep=1pt,parsep=0pt]
\item \textbf{problem\_io / step 80.7 (MD\_NVT function\_header)} \,\textit{scientific: Spec$\leftrightarrow$gold contradiction} \,$\cdot$\, too-strict/wrong\\
{\footnotesize The spec described a Berendsen thermostat plus barostat (NPT ensemble) and claimed a modified box length is returned, directly contradicting the problem's actual algorithm (Andersen NVT, no barostat) and the true five-element return; a solver following the original spec would implement the wrong ensemble and wrong return contract.}
\item \textbf{step 80.2 (E\_ij) and step 80.4 (f\_ij) function headers} \,\textit{surface: Interface / return-contract mismatch} \,$\cdot$\, neutral-cleanup\\
{\footnotesize The E\_ij and f\_ij docstrings mentioned a truncated-and-shifted Yukawa potential as an optional component, but no Yukawa parameters exist anywhere in the task; the spurious mention misleads solvers without affecting grading, since the gold tests only exercise Lennard-Jones.}
\item \textbf{step 80.3 (background), step 80.4 (f\_ij r-type), step 80.5 (forces Returns)} \,\textit{scientific: Unspecified convention} \,$\cdot$\, too-strict/wrong\\
{\footnotesize The background limited energy to `two atoms next to each other' instead of all distinct pairs (i<j) at minimum-image distance; f\_ij's argument r was typed as a scalar float when the gold requires the signed 3D displacement vector r\_j - r\_i; and the Newton's-third-law force accumulation sign convention was left implicit, so correct vectorial implementations could be graded wrong.}
\item \textbf{step 80.5 (forces) test\_cases} \,\textit{surface: Trivially broken test} \,$\cdot$\, too-strict/wrong\\
{\footnotesize The forces() test declared N=2 but supplied a three-row positions array; a correct implementation iterating over N=2 would silently ignore the third particle, making the test input self-contradictory and the expected output physically wrong.}
\item \textbf{step 80.7 (MD\_NVT) test\_cases and general\_tests} \,\textit{scientific: RNG-dependent grading} \,$\cdot$\, too-strict/wrong\\
{\footnotesize N=200 conflicts with initialize\_fcc, which returns 6\textasciicircum{}3=216 lattice sites, so the declared particle count mismatches the actual positions array; additionally, the Andersen thermostat makes stochastic collision draws, so without a fixed RNG seed the temperature and energy assertions are non-reproducible across runs.}
\end{itemize}

\end{document}